\documentclass[11pt,times]{article}
\usepackage{amscd,amsfonts,amsmath,amstext,enumerate,float,amsthm,amssymb,epsfig,multirow,color,latexsym,mathrsfs,tocloft,tikz,tkz-fct,listofsymbols,tabu}
\usepackage{algcompatible}
\usepackage{bbm}
\usepackage{booktabs}
\usepackage{subcaption}
\usepackage{graphicx}
\usepackage[titletoc]{appendix}
\graphicspath{ {Figures/} }
\usetikzlibrary{decorations.pathreplacing}
\usepackage[letterpaper,left=3.81cm,top=2.54cm,bottom=2.54cm,right=2.54cm]{geometry}
\usepackage[nottoc]{tocbibind}

\usepackage{appendix}
\usepackage{chngcntr}
\usepackage{etoolbox}

\usepackage{tikz,tikz-cd}
\usepackage{tkz-euclide}
\usetkzobj{all}
\usetikzlibrary{matrix,shapes,arrows}
\usetikzlibrary{positioning}
\usepackage{smartdiagram}
\AtBeginEnvironment{subappendices}{%
	\chapter*{Chapter Appendix}
	\addcontentsline{toc}{chapter}{Appendices}
	\counterwithin{figure}{section}
	\counterwithin{table}{section}
}

\usepackage{graphicx}
\usepackage{sidecap}
\usepackage{float}
\usepackage{enumerate,mdwlist}
\usepackage{epsfig}
\usepackage[english]{babel}
\usepackage[bbgreekl]{mathbbol}
\usepackage{amsfonts,amsmath,enumerate}
\usepackage{amscd}
\usepackage{amsthm}
\usepackage{mleftright}
\DeclareSymbolFontAlphabet{\mathbb}{AMSb}
\DeclareSymbolFontAlphabet{\mathbbl}{bbold}
\usepackage{colortbl}

\usepackage{subcaption}
\usepackage{amsmath,amsfonts,amssymb,mathrsfs}\usepackage{bm}
\usepackage{latexsym}
\usepackage{mleftright}
\usepackage{amssymb}
\usepackage{color}
\usepackage{xifthen}

\usepackage[numbers]{natbib}%
\usepackage[titletoc]{appendix}

\usepackage{footnote}

\newlength{\oldparindent}
\usepackage{fancyhdr}
\usepackage{hyperref} 
\hypersetup{
	colorlinks = true,
	linkcolor = blue,
	anchorcolor = blue,
	citecolor = blue,
	filecolor = blue,
	urlcolor = blue
}

\usepackage[ruled,vlined,resetcount,algosection]{algorithm2e}
\usepackage{makeidx}
\usepackage{hyperref}
\usepackage{amsmath}
\usepackage{amsthm}
\usepackage{amsfonts}
\usepackage{amssymb}
\usepackage{mathrsfs}
\usepackage{graphicx}
\usepackage{tikz,tikz-cd}
\usepackage{tkz-euclide}
\usetkzobj{all}
\usetikzlibrary{matrix}
\usepackage{smartdiagram}

\makeatletter
\newcommand{\superimpose}[2]{%
	{\ooalign{$#1\@firstoftwo#2$\cr\hfil$#1\@secondoftwo#2$\hfil\cr}}}
\makeatother

\newcommand{\rr}{{\mathbb{R}}}
\newcommand{\rrflex}[1]{{\ensuremath{\rr^{#1}
}}}
\newcommand{\rrD}{{\rrflex{D}}}
\newcommand{\rrd}{{\rrflex{d}}}

\newcommand{\xxx}{\mathcal{X}}

\newcommand{\uuu}{\mathcal{U}}

\newcommand{\cc}{{\mathbb{C}}}

\newcommand{\pp}{{\mathbb{P}}}

\newcommand{\mmm}{{\mathscr{M}}}

\newcommand{\hhh}{{\mathscr{H}}}

\newcommand{\fff}{{\mathscr{F}}}

\newcommand{\thetareg}[1]{{\ensuremath{
			\theta_t^{\mathfrak{ext}}
}}}
\newcommand{\thetarreg}[1]{{\ensuremath{
			\thetar{t}%
}}}

\newcommand{\thetar}[1]{{
		\ensuremath{
			\theta_{#1}^{\mathfrak{ext}}
		}
}}

\newcommand{\GamLim}[1]{{
		\ensuremath{
			\underset{#1}{
				\operatorname{\Gamma-lim}\,	
			}
		}
}}

\makeatletter
\newcommand*\bigcdot{\mathpalette\bigcdot@{.5}}
\newcommand*\bigcdot@[2]{\mathbin{\vcenter{\hbox{\scalebox{#2}{$\m@th#1\bullet$}}}}}
\makeatother

\makeatletter
\def\@chapter[#1]#2{\ifnum \c@secnumdepth >\m@ne
	\refstepcounter{chapter}%
	\typeout{\@chapapp\space\thechapter.}%
	\addcontentsline{toc}{chapter}%
	{\protect\numberline{\thechapter}\string\hypertarget{chap\thechapter}{#1}}%
	\else
	\addcontentsline{toc}{chapter}{#1}%
	\fi
	\chaptermark{#1}%
	\addtocontents{lof}{\protect\addvspace{10\p@}}%
	\addtocontents{lot}{\protect\addvspace{10\p@}}%
	\if@twocolumn
	\@topnewpage[\@makechapterhead{#2}]%
	\else
	\@makechapterhead{#2}%
	\@afterheading
	\fi}
\def\@makechapterhead#1{%
	\vspace*{50\p@}%
	{\parindent \z@ \raggedright \normalfont
		\ifnum \c@secnumdepth >\m@ne
		\huge\bfseries \@chapapp\space \thechapter
		\par\nobreak
		\vskip 20\p@
		\fi
		\interlinepenalty\@M
		\Huge \bfseries \hyperlink{chap\thechapter}{#1}\par\nobreak
		\vskip 40\p@
}}
\makeatother

\newcommand{\xireg}[1]{{\ensuremath{
			\xi_t^{\mathfrak{r}}
}}}
\newcommand{\xirreg}[1]{{\ensuremath{
			\xir{t}%
}}}

\newcommand{\xir}[1]{{
		\ensuremath{
			\xi_{#1}^{\mathfrak{r}}
		}
}}

\newtheoremstyle{dotless}{}{}{\itshape}{}{\bfseries}{}{ }{}
\theoremstyle{dotless}

\newtheorem{defn}{Definition}[section]

\theoremstyle{plain}
\theoremstyle{definition}
\newtheorem{prop}[defn]{Proposition}%
\newtheorem{thrm}[defn]{Theorem}%
\newtheorem{remark}[defn]{Remark}%
\newtheorem{setup}[defn]{Notation}
\newtheorem{ass}[defn]{Assumption}

\newcommand{\describeContent}[1]{%
	\begingroup%
	\let\thefootnote\relax%
	\footnotetext{#1}%
	\endgroup%
}

\NewDocumentCommand\argmin{o}{{\operatorname{argmin}\IfValueT{#1}{_{{#1}}}}}
\NewDocumentCommand\AF{o}{\operatorname{AF}\IfValueT{#1}{
			{
				\left({#1}\right)
			}
		}}

\newcommand{\ddd}{{\mathcal{D}}}
\newcommand{\vv}{{\Delta}}
\NewDocumentCommand{\ee}{moo}{
	{
		\mathbb{E}_{\IfValueF{#3}{\pp}\IfValueT{#3}{{#3}}}\left[
		{#1}
		\IfValueT{#2}{\mid {#2}}
		\right]
	}
}
\NewDocumentCommand\AIR{moo}{
{
		\mathbb{A}%
		\IfValueT{#3}{_{#3}}\IfValueF{#2}{_{S}}
		\left(
		{#1}\mid\IfValueT{#2}{{#2}}\IfValueF{#2}{{\hhh}}
		\right)
}
		}

\usepackage{fancyhdr}
\NewDocumentCommand{\tr}{o}{{\operatorname{tr}\IfValueT{#1}{\left({#1}\right)}}}
\begin{document}
\author{Anastasis Kratsios\thanks{Department of Mathematics, Eidgen\"{o}ssische Technische Hochschule Z\"{u}rich, HG G 32.3, R\"{a}mistrasse 101, 8092 Z\"{u}rich.  email: \textit{anastasis.kratsios@math.ethz.ch}}\,\, Cody Hyndman\thanks{Department of Mathematics and Statistics, Concordia University, 1455 Boulevard de Maisonneuve Ouest, Montr\'{e}al, Qu\'{e}bec, Canada H3G 1M8. email: \textit{cody.hyndman@concordia.ca}}
}
\title{\Large \bf Deep Learning in a Generalized HJM-type Framework Through Arbitrage-Free Regularization} %
\lhead{A. Kratsios, C. Hyndman
}
\chead{\small Deep AF Regularization}
\rhead{\small Dec 5$^{th}$ 2019}%

\date{%
	December 5$^{th}$ 2019 %
}

\maketitle

\begin{abstract}
We introduce a regularization approach to arbitrage-free factor-model selection.   The considered model selection problem seeks to learn the closest arbitrage-free HJM-type model to any prespecified factor-model.  An asymptotic solution to this, a priori computationally intractable, problem is represented as the limit of a 1-parameter family of optimizers to computationally tractable model selection tasks.  Each of these simplified model-selection tasks seeks to learn the most similar model, to the prescribed factor-model, subject to a penalty detecting when the reference measure is a local martingale-measure for the entire underlying financial market. A simple expression for the penalty terms is obtained in the bond market withing the affine-term structure setting, and it is used to formulate a deep-learning approach to arbitrage-free affine term-structure modelling.   Numerical implementations are also performed to evaluate the performance in the bond market.  
\end{abstract}

	\noindent
{\itshape Keywords:}
Arbitrage-Free Regularization, Consistent HJM Models, $\Gamma$-Convergence, Bond Pricing, Model Selection, Deep Learning.

\noindent
\let\thefootnote\relax\footnotetext{This research was supported by the ETH Z\"{u}rich Foundation as well as by the Natural Sciences and Engineering Research Council of Canada (NSERC).  The authors thank Alina Stancu for many helpful discussions.}
\section{Introduction}\label{s_Intro}
The price of a {zero-coupon bond} has been modelled as a function of the instantaneous interest rate in effect at a given time; this approach providing many useful results within the affine and quadratic term-structure setting of \cite{Duffie,vasicek1977equilibrium,ahn2002quadratic}.  However, the difficulty of adequately calibrating the model to all the available term-structure of interest data had lead \cite{heath1992bond} and latter \cite{TappeFilLevyHJM,malliavin2007non,jakubowski2007exponential} to instead describe the evolution of zero-coupon bond prices indirectly by viewing them as a function of a latent variable, the instantaneous {forward-rate curve}.  This stochastically evolving curve describes all future interest rates as viewed from the current time.  

The impracticality of the infinite-dimensional nature of forward-rate curve models is typically surmounted by turning to a factor-model approach, for example, see \cite{nelson1987parsimonious,waggoner1997spline,ludvigson2009factor}.  However, it has since been shown in \cite{filipovic2000exponential} that a large number of forward-rate curve factor-models allow for arbitrage in the bond market.  Though some authors, such as \cite{AFNS}, have found perturbations of specific widely used models that bypass this issue, it is typically accepted that (finite) factor-models for the forward-rate curve fail to provide a theoretically satisfactory view of the bond market.  

The objective of this paper is to introduce a general framework for optimally solving the problem of learning the most similar arbitrage-free factor model to a prespecified factor model.  The method works by analogy with traditional finance where, given a complete financial market, the risk-neutral price of a financial asset is found by perturbing it's price process by a multiplicative martingale factor until it becomes a local martingale for the reference measure.  Similarly, the method introduced in this paper perturbs a given factor-model for the forward-rate curve until the induced bond prices simultaneously become local-martingales under the reference probability measure.  

In our setting, the perturbations of the factor-model for the forward-rate curve are achieved by searching a prespecified {hypothesis class} of alternative models for the most similar factor-model for the forward-rate curve for which all the bond price processes becomes local-martingales under the reference probability measure.  This search is formalized by minimizing a loss-function measuring the distance of the alternative model to the original forward-rate curve model, with the additional constraint that the resulting bond processes define local-martingales.  

The second innovation of this paper lies in a computationally tractable relaxation of this optimization problem whose optimizers are shown to, asymptotically, converge to the optimizers of the formerly described problem.  The flexibility of deep feedforward neural networks, as established in \cite{hornik1990universal,Cybenko,hornik1991approximation}, makes them an ideal computationally tractable hypothesis class of alternative factor-models for the forward-rate curve.  Implementations of our arbitrage-free regularization theory using deep feedforward neural networks are also considered in this paper.  
Just as the modelling framework of \cite{heath1992bond} is not limited to the bond market and has since been extended to a variety of asset classes in \cite{dupire1994pricing,rita2002design,carmona2007hjm,kallsen2015heath,kratsios2018arbitrage}, our approach is not bond-specific and the problem is treated in full generality within this paper.  
\subsection{Background}\label{ss_Background}
Before formalizing and solving the arbitrage-free regularization problem, some relevant background is briefly discussed.  These topics include related aspects of the functional It\^{o} calculus introduced in \cite{dupire2009functional} and developed in \cite{fournie2010functional}, as well as pertinent stochastic differential geometric considerations; as developed in \cite{elworthy1982stochastic}.  Some elements from arbitrage-theory are also discussed concisely.  All stochastic processes described in this paper are defined on a common stochastic base $\left(\Omega,\mathcal{F},\left\{\mathcal{F}_{t}\right\}_t,\pp\right)$.  
\subsubsection{{Functional It\^{o} Calculus}}
The functional It\^{o} calculus of \cite{dupire2009functional} and \cite{fournie2010functional}, has found many applications in mathematical finance.  Applications range from, but are not limited to, computational methods for the Greeks of path-dependent options in \cite{jazaerli2017functional} to portfolio theory in \cite{pang2015application}.  The basic concept relies on non-anticipative and path-dependent extensions of the time and spatial derivative operators.  Briefly, both these extensions are constructed by lengthening any suitable c\'{a}dl\'{a}g path, by artificially extending its endpoint either vertically or horizontally.  Analogously to classical calculus, the limiting ratio between the difference of original path with its artificial extension, with the length of time the path was artificially extended approaches $0$.  In the case where vertical extensions of the path were chosen, one finds a notion of derivative extending the classical derivative, aptly called the {spatial derivative} and denoted by $\vv$.  Likewise, in the case where the horizontal extensions were considered, a path-dependent extension of the derivative for the time parameter is obtained; it is called the {time derivative} and is denoted by $\ddd$.  

Using the time and spatial derivatives a second-order calculus defined on functionals of a c\'{a}dl\'{a}g path, and thus extending the deterministic It\^{o} calculus of \cite{follmer1981calcul}, is developed in \cite{dupire2009functional}.  However, unlike the classical It\^{o} calculus, which is defined for $C^{1,2}(I\times \rrd)$-functions it is shown in \cite{fournie2010functional} that it is not enough for a path-dependent functional to admit a continuous time derivative and two continuous spatial derivatives; where $I\triangleq [0,\infty)$  and continuity is defined with respect to the following metric 
$$
d
(x_{t},y_{s})\triangleq 
\sup_{u \in [0,s]}\|
x_{t,s-t}(u) - y_u 
\| + |s-t|
;
$$
where $t\leq s$ and $x_{t,s-t}(u)$ is the extension of the path which agrees with $x_t$ up to time $t$ and then maintains the constant value $x_t$ until time $s$, evaluated at time $u \in [0,s]$.  For this calculus, it is additionally required that functionals of the path be bounded on segments of the paths taking values in any non-empty compact subset of $\rrd$.  By analogy with the classical calculus, the collection of all non-anticipative path-dependent functionals satisfying these three requirements is denoted by $\cc^{1,2}_b$ and strictly contains $C^{1,2}(\rr)$ as a subset. The collection $\cc^{1,2}_b$, therefore, describes all path-dependent functionals on which this second-order path-dependent calculus is well-defined.  For more details on the functional It\^{o} calculus, the reader is referred to \cite{fournie2010functional,IBPandFICalculus,MartingaleRepFunctionalItoFournieRama}.  
%
%
%

Next, some relevant notions from stochastic differential geometry will be briefly discussed.  
\subsubsection{Stochastic Differential Geometry}
Numerous authors, such as \cite{brigo2002lognormal,brody2004chaos,Henryboyasymptotic,GeoFiltering,gidea2017topological,kratsios2017geometric}, have exploited geometric modelling approaches in mathematical finance, with applications ranging from volatility surface modelling to the term-structure of interest.  In order to allow the theory developed here to be compatible with these methods, a stochastic differential geometric modelling approach is considered.  Some key aspects of this theory, pioneered in \cite{elworthy1982stochastic}, will are be summarized to fix notation.

Given a $d$-dimensional Riemannian manifold $(\mmm,g_t)$ with a (potentially time-varying) Riemannian metric $g_t$ and an $\rrd$-valued diffusion process defined as the unique strong solution\footnote[1]{Conditions for existence and uniqueness of a strong solution are discussed in the main body of the paper.} to the SDE
\begin{equation}
b_t = b_0 + \int_0^t \mu(s,b_s)ds + \int_0^t\sigma(s,b_s)dW_s
\label{eq_stoch_anti_development_factor_process}
,
\end{equation}
one may construct, in a unique manner, an $\mmm$-valued semi-martingale which is compatible with the geometry described by $g_t$; where here $b_0\in \rrd$.  This is typically interpreted as rolling $b_t$ tangentially along $\mmm$ without any sudden movements.  This rolling procedure is made possible by first continuously associating each $b_t(\omega)$ to an orthonormal basis $U_t(\omega)$ lying tangent to $\mmm$ and then projecting $U_t(\omega)$ onto $\mmm$.  The process $U_t$ lies on the larger manifold $\mathscr{O}(\mmm)$ obtained by gluing of all possible orthonormal basises tangentially to each point on $\mmm$.  It is shown in \cite{elworthy1982stochastic} (and in \cite{GuoTimeDep} for the time-dependent case) that there exists exactly one way, given an initial choice of $U_0$ and $b_0$, to project the tangential process $U_t$ down to $\mmm$ in such a way that the resulting $\mmm$-valued process is compatible with the geometry described by $g_t$.  The resulting process, which will be denoted by $\beta_t$, is called the $g_t$-horizontal stochastic development of $b_t$ along $\mmm$ (with initial frame $U_0$).  The terminology stems from the approach to differential geometry developed in \cite{ehresmann1948connexions}.  

It is shown in \cite{elworthy1982stochastic,GuoTimeDep} that there is a well-defined second-order calculus for $\beta_t$ which extends the traditional It\^{o}-calculus on $\rrd$.  Moreover, for $\rr$-valued functions $\phi$ which are once-differentiable in time and twice differentiable in their spatial variable, the following It\^{o}-formula holds, write in local-coordinates
\begin{equation}
\begin{aligned}
\phi(t,\beta_t) 
	= &
\int_0^t 
\frac{\partial \phi}{\partial t}(s,\beta_s) 
ds
+
\int_0^t
\sum_{i=1}^d \frac{\partial \phi}{\partial\beta^i}(s,\beta_s)db_s^i
\\
+ &
\int_0^t
\frac1{2}
\sum_{i,j=1}^{d}
\left(
		\frac{\partial^2 \phi}{\partial\beta^i\partial\beta^j}
												(s,\beta_s)
	-
		\sum_{k=1}^{d}\Gamma^k_{i,j}(t)
			\frac{\partial^k \phi}{\partial\beta^k}
												(s,\beta_s)
\right)
d[b]^{i,j}_s
ds
,
\end{aligned}
\label{eq_Ito_formula_manifolds}
\end{equation}
where $\Gamma_{i,j}^k$ are the Christoffel symbols of $(\mmm,g_t)$ at-time $t$; the symbols $\Gamma_{i,j}^k$, compare the local curvature of $(\mmm,g_t)$ to that of Euclidean space.  For example, when $(\mmm,g_t)$ is itself Euclidean space, then $\Gamma_{i,j}^k(t)=0$, thus~\eqref{eq_Ito_formula_manifolds} reduces to the usual It\^{o}-formula.  In this case, it can be deduced from~\eqref{eq_Ito_formula_manifolds} that $\beta_t=b_t$ (up to the choice of initial data).  

Next some background on arbitrage theory in large financial markets is discussed.  
\subsubsection{Arbitrage-Theory}
The efficient market hypothesis, introduced in \cite{bachelier1900theorie}, states that the typical market participant cannot earn a risk-less profit.  The efficient market hypothesis has found several mathematical formulations, as summarized in \cite{fontana2014note,fontana2015weak}.  The most commonly used form is No Free Lunch with Vanishing Risk (NFLVR) as formulated in the sequence of papers \cite{FTOAP,FTOAP1,FTOAPHilbertDiscrete} which builds on the ideas of \cite{KrepsFTOAPGen}.  Essentially, in the case of locally bounded processes, NFLVR expresses the non-existence of arbitrage-strategies as 
the existence of an equivalent local martingale measure (ELMM); that is, a probability measure which is equivalent to the reference probability measure and which simultaneously makes the price process of each market asset a local martingale.

However, mathematically bond markets are unlike traditional financial markets in that they are comprised of an uncountable number of assets, one for each potential maturity; thus, the results of \cite{FTOAP,FTOAP1} no longer apply since their formulation requires that only a finite number of assets be tradeable.  Instead, in the setting of such a \textit{large financial market}, a satisfactory and economically meaningful no-arbitrage condition is obtained in \cite{AsymptoticNAKleinSchachermayer,NewPerspectiveFTOAPLarge} by considering strategies which can be described by limits of classical strategies written on a finite number of market assets.  It is shown in \cite{NewPerspectiveFTOAPLarge}, that when each asset in the market is locally bounded, as in \cite{FTOAP,klein2013roll}, then the no-arbitrage condition derived in \cite{NewPerspectiveFTOAPLarge} reduces to the existence of an equivalent local martingale measure.  However, if the local-boundedness assumption is dropped, then the existence of an equivalent local martingale measure remains sufficient for precluding no-arbitrage but it no longer necessary.  This last point is important throughout our analysis.  
%
%
The central problem of focus in this paper is now formulated.  
\subsection{The Arbitrage-Free Regularization Problem}\label{ss_Problem_Statement}
The paper is concerned with the modelling of a large financial market $\{X_t(u)\}_{u \in \uuu}$, indexed by a non-empty Borel subset $\uuu\subseteq \rrD$; were $D$ is a positive integer.  In the case of the bond market, following \cite{musiela1993stochastic,bjork1999interest,filipovic2001consistency,FilTapTeich}, $\uuu=[0,\infty)$ represents the collection of all possible \textit{times to maturity} and $X_t(u)$ represents the time $t$ price of a zero-coupon bond with maturity $T\triangleq u+t$.  As observed in \cite{musiela1993stochastic}, the choice of parameterizing with respect to time to maturity removes the dependence on time in $\uuu$.  

For each $u \in \uuu$, the process $X_t(u)$ will be driven by an unobservable latent factor process; in the case of the bond market, this latent process will be the forward-rate curve.  The relationship between $X_t(u)$ and the latent process will be expressed by
\begin{equation}
\begin{aligned}
X_t(u)\triangleq& S_t\left(
\phi_t^u,[\phi^u]_t ;u
\right)
&\qquad \phi_t^u\triangleq \phi(t,\beta_{t},u)
;
\end{aligned}
\label{eq_definition_market}
\end{equation}
where $\{S_t(\cdot,\cdot;u)\}_{u \in \uuu}$ is a family of path-dependent functionals encoding the latent process into the asset price $X_t(u)$, $\phi_t^u$ is the factor-model for the latent process, and $\beta_t$ are the $\mmm$-valued stochastic factors driving the latent process; where following \cite{fournie2010functional}, $S_t$ will be allowed to depend on the quadratic-variation of the factor process $\phi(t,\beta_t,u)$.  By the Nash Embedding Theorem of \cite{nash1956imbedding}, there is no loss of generality in assuming that $\mmm\subseteq \rr^N$ for a sufficiently large positive integer $N$ and this assumption will also be maintained for convenience.

In the case of the bond market, $S_t$ will be the map taking a forward-rate curve to the price of a zero-coupon, as defined by
\begin{equation}
S_t\left(
\phi_t^u,[\phi^u]_t ;u
\right)
\triangleq \exp\left(
\int_t^{u+t} \phi(t,\beta_t,v)dv
\right)
\label{eq_structure_map_bond_case}
.
\end{equation}

In general, $S_t$ will be allowed to depend on the path of $X_t(u)$; thus $S_t$ will be a path-dependent functional of regularity $\cc^{1,2}_b$ in the sense of \cite{fournie2010functional}.  However, as in the bond market, if $S_t$ depends only on the current value of $X_t$ then the requirement that $S_t$ be of class $\cc^{1,2}_b$, in the sense of \cite{fournie2010functional}, is equivalent to it being of regularity $C^{1,2}(I\times \rrd)$ in the classical sense; where $I\triangleq [0,\infty)$.  Whence, the classical It\^{o}-calculus would apply to $S_t$.  

Analogously to \cite{bjork1999interest,filipovic2000finite,de2017finite}, the factor-model $\phi$ for the latent process will always be suitably integrable and suitably differentiable.  Specifically, $\phi$ will belong to a Banach subspace $\xxx$ of $L_{w\otimes \mu}^p\left(I\times \mmm\times \uuu\right)$ which can be continuously embedded within the Fr\'{e}chet space $C^{1,2,2}(I\times \mmm\times \uuu)$; where $\nu$ is a Borel probability measure supported on $I$, $\mu$ is a Borel probability measure supported on $\mmm\times \uuu$, and both $\nu$ and $\mu$ are equivalent to the corresponding Lebesgue measures restricted to their supports.  Here, $1\leq p < \infty$ is kept fixed.  
%

A recurring example from the bond modelling literature is the Nelson-Siegel model of \cite{nelson1987parsimonious}, as described in \cite{diebold2006forecasting,YieldDbold}, which expresses the forward-rate curve as a function of its level, slope, and curvature through the factor-model.  The Nelson-Siegel family are part of a larger class of affine term-structure models, in which, at any given time, the forward-rate curve is described in terms of a set of market factors as
\begin{equation}
\begin{aligned}
\varphi(t,\beta,u)\triangleq \varphi_0(u+t)+\sum_{i=1}^{d} \beta^i\varphi_i(u+t)
,
\end{aligned}
\label{eq_ATS}
\end{equation}
where $d$ is a positive integer and $\varphi_i \in \xxx$ and $\varphi_0$ is a forward-rate curve typically calibrated to the data available at time $t=0$.  However, as shown in \cite{filipovic2001consistency}, the Nelson-Siegel model is typically not arbitrage-free therefore we would like to learn the closest arbitrage-free factor-model, driven by the same stochastic factors.  

%
%


  The central problem of this paper is to identify an optimal factor-model within a given hypothesis class $\hhh\subseteq \xxx$, of plausible alternative models.  This question is formalized through the following optimization problem
\begin{equation}
\begin{aligned}
&\argmin[\phi \in \hhh] \,
\ell\left(
\varphi- \phi
\right)
\\
&\mbox{subject to:} \,  S_t(\phi_t^u,[\phi^u]_t ;u) \mbox{ is a $\pp$-local martingale for all $u \in \uuu$}
;
\end{aligned}
\tag{AFProj}
\label{AF}
\end{equation}  
where $\hhh$ is required to contain the (naive) factor-model $\phi$ and $\ell:\xxx\rightarrow [0,\infty)$ is a lower semi-continuous and coercive loss function.  For example, $\ell$ may be taken to be the norm on $\xxx$.  Geometrically,~\ref{AF} describes a projection of $\varphi$ onto the (possibly non-convex) subset of $\hhh$ of factor-models making each $S_t(\phi_t^u,[\phi^u]_t ;u)$ into a $\pp$-local martingale for every $u\in \uuu$.  

In general, the problem described by~\eqref{AF} may be challenging to implement as projections onto non-convex sets are not well-behaved.  Thus, in analogy with regularization literature such as~\cite{hoerl1970ridge,hastie2015statistical,ENETZou2005regularization}, in statistical learning theory, one may instead consider the following relaxation of the~\eqref{AF} since it is more amenable to numerical implementation
\begin{equation}
\argmin[\phi \in \hhh] \,
\ell\left(
\varphi - \phi
\right)
	+
\AF^{\lambda}(\phi)
\tag{AFReg}
\label{AFReg}
;
\end{equation}
where $\{\AF^{\lambda}\}_{2\leq \lambda < \infty}$ is a family of functions from $\hhh$ to $[0,\infty]$ taking value $0$ if each \\
$S_t(\phi_t^u,[\phi^u]_t ;u)$ is a $\pp$-local martingale simultaneously for every value of $u$ and where $\lambda$ is a meta-parameter determining the amount of emphasis placed on the penalizing factor-models which fail to meet this requirement. ~\eqref{AFReg} will be
 called the \textit{arbitrage-free regularization} problem.  

Numerical solutions to~\eqref{AFReg} have been considered in \cite{kratsios2018arbitrage} within in continuous time framework for the bond-market and discrete-time analogues of~\eqref{AFReg} have been implemented by \cite{PelgerLatent,PelgerCrosssectional,PelgerDeep} for a portfolio of stocks.  In both these situations, it is numerically understood that~\eqref{AFReg} provides a suitable estimate of the solution to~\eqref{AF}.  The second core objective of this paper is to establish this fact.  

Thus, the central mathematical program of this paper may be outlined as follows.  First, a family of non-negative penalty functions $\AF^{\lambda}$ taking value $0$ when the evaluated factor-model defines a family of $\pp$-local martingales will be constructed.  Second, it will be shown, using the theory of $\Gamma$-convergence developed in \cite{de1980gamma}, that the optimizers of~\eqref{AFReg} converge to an optimizer of~\eqref{AF}.  This is the content of Section~\ref{s_Full_problem}.  
Once the problem has been addressed in full generality, Section~\ref{s_AF_Reg_FRC} returns to the motivating bond example.  Wherein, the flexibility of deep feedforward neural networks are used to approximate~\eqref{AFReg} and comparisons are drawn to other methods.  

For convenience, the precise assumptions and notation will now be summarized.
\begin{setup}\label{notation_summary}
	The following notation will be maintained throughout this paper.    
	\begin{enumerate}[(i)]
		\item $I\triangleq [0,\infty)$.
		\item $(\mmm,g_t)$ is a Riemannian sub-manifold of $(\rr^{N},g_{EUC})$ with time-dependent connection, where $g_{EUC}$ is the usual Riemannian metric on $\rr^{N}$.  An initial frame $U_0$ will always be fixed.  
		\item $\mu$ is a 
		 Borel probability measure on $I\times \uuu$ (with respect to its relative topology) which is equivalent to the Lebesgue measure on its support,
		 \item $C^{1,2,2}(I\times \mmm\times \uuu)$ is the space of functions from $I\times \mmm\times \uuu$ to $\rr$ admitting one derivative in the first input and two derivatives in its other inputs; it will be topologized by the following family of semi-norms
		 $$
		 p_{\gamma,\alpha,\delta,K}(\phi)
		 \triangleq 
		 \sup_{|\gamma|\leq 1, |\alpha|\leq 2,|\delta|\leq 2} 
		 \sup_{(t,\beta,u) \in K} 
		 \left|
		 \frac{\partial^{\gamma}}{\partial s^{\gamma}}
		 \frac{\partial^{|\alpha|}}{\partial \beta_i^{\alpha_i}\dots \beta_j^{\alpha_j}}
		 \frac{
		 	\partial^{|\delta|}
		 }{
		 	\partial u_{\tilde{i}}^{\delta_l}\dots \partial u_{\tilde{j}}^{\delta_k}
		 }
		 \phi\left(t,\beta,u\right)
		 \right|
		 ;
		 $$
		 where $K$ is a non-empty compact subset of $I \times \mmm\times \uuu$ and $\gamma,\alpha$, and $\delta$ are (multi-)indices.
	\end{enumerate}
\end{setup}
\begin{ass}\label{assumptions_summary}
	The following assumptions will be maintained throughout this paper.    
	\begin{enumerate}[(i)]
		\item $\beta_t$ is a $\mathcal{F}_t$-adapted, $g_t$-horizontal semi-martingale with fixed initial frame $\Xi_0\in \mathcal{O}(\mmm)$ and stochastic anti-development $b_t$ which is the unique strong solution to the $\rrd$-valued SDE~\eqref{eq_stoch_anti_development_factor_process}; where $W_t$ is an $\rrd$-valued Brownian motion, the components $\mu^i:\rr^{1+d}\rightarrow \rr$ are globally $K$-Lipschitz, the components $\left(\sigma^{i,j}:\rr^{1+d}\rightarrow \rr^{d\times d} \right)_{i,j=1}^d$ are globally $\tilde{K}$-Lipschitz, where $K,\tilde{K}>0$.
		\item $\phi\in \xxx$, where $\xxx$ is a Banach subspace of $L_{\nu\otimes \nu}^p\left(I\times \mmm\times \uuu\right)$ which admits a continuous embedding into $C^{1,2,2}(I\times \mmm\times \uuu)$.  Furthermore, $\xxx$ will always be viewed as continuously embedded within this space.
		\item For every $u \in \uuu$, $\{S_t(\cdot,\cdot;u)\}_{t \in [0,\infty)}$ is a non-anticipative functional in $\mathbb{C}_b^{1,2}$ verifying the following "predictable-dependence" condition, of \cite{fournie2010functional},
		\begin{equation*}
		\begin{aligned}
		S_t(x_t,x_t ;u)&=S_t(x_t,x_{t_-} ;u)
		&\qquad
		\left(\forall t \in [0,\infty)\right)
		\, \left(\forall (x,v)\in D([0,t];\rrd)\times D([0,t];S_+^d)\right)
		,
		\end{aligned}
		\end{equation*}
		where $S_+^d$ is the set of $d\times d$-dimensional positive semi-definite matrices with real-coefficients,
		\item The hypothesis class $\hhh\subseteq \xxx$ is a non-empty and unbounded.  
	\end{enumerate}
\end{ass}
%
%
%
The central problem of the paper will now be addressed in full generality before turning to applications in term-structure models.  
\section{{A General Solution to~\eqref{AFReg} and~\eqref{AF}}}\label{s_Full_problem}
In this section, we show the asymptotic equivalence of problems~\eqref{AF} and~\eqref{AFReg} for general asset classes.  This first requires the construction of the penalty term $\AF^{\lambda}$, measuring how far a given factor-model is from being a $\pp$-local martingale.  The construction of $\AF^{\lambda}$ is made in two steps, first a drift condition ensuring that each $\{X_t(u)\}_{u \in \uuu}$ is simultaneously a $\pp$-local martingale, this generalizes the drift condition of \cite{heath1992bond} and it provides an analogue to the consistency condition of \cite{filipovic2004geometry}.  In the second step, the drift condition will then be used to build the penalty term in~\eqref{AFReg}.  Subsequently, the optimizers of~\eqref{AFReg} will be used to asymptotically solve~\eqref{AF}.  
\begin{prop}[Drift Condition]\label{prop_NA_characterization}
	Let $\phi,S,\beta$ be as above.  Then $X_t(u)$ is a $\pp$-local-martingale, for each $u \in \uuu$ simultaneously, if and only if
\begin{equation}
\hspace*{-.5em}
\begin{aligned}
-\ddd S_s(\phi^u_s,[\phi^u]_s;u)  = &
\vv S_s(\phi^u_s,[\phi^u]_s;u) 
\left[
\frac{\partial \phi}{\partial t}(s,\beta_s,u) 
+
\sum_{i=1}^d \frac{\partial \phi}{\partial\beta^i}(s,\beta_s,u)\mu^i(s,b_s)
\right.\\
+ &
\left.
\frac1{2}
\sum_{i,j=1}^{d}
\left(
\frac{\partial^2 \phi}{\partial\beta^i\partial\beta^j}
(s,\beta_s,u)
-
\sum_{k=1}^{d}\Gamma^k_{i,j}(t)
\frac{\partial^k \phi}{\partial\beta^k}
(s,\beta_s,u)
\right)
\sigma^i(s,b_s)\sigma^j(s,b_s)
\right]
\\
+ &
\frac1{2} \tr[
\vv^2 S_s(\phi^u_s,[\phi^u]_s;u) 
]\left(
\sum_{i=1}^d \frac{\partial \phi}{\partial\beta^i}(s,\beta_s,u)\sigma^i(s,b_s)
\right)^2
.
\end{aligned}
\label{eq_thrm_NA_condition}
\end{equation}
	is satisfied for every $t \in [0,\infty)$ and every $u \in \uuu$; $\pp$-a.s.    
\end{prop}
\begin{proof}
For legibility, for each $u \in \uuu$, we represent the process $\phi(t,\beta_t,u)$ by 
$$
\phi_t^u = \phi_0^u + \int_0^t \alpha^u(s,\phi_s^u)ds + \int_0^t \gamma^u(s,\phi_s^u)dW_s
.
$$
By the Functional It\^{o} Formula, \citep[Theorem 4.1]{MartingaleRepFunctionalItoFournieRama}, if follows that, for every $u \in \uuu$
\begin{equation}
\begin{aligned}
S_t(\phi^u_t,[\phi^u]_t;u) = & S_t(0,\phi^u_0,[\phi^u]_0;u) \\
	+ &
\int_0^t \left[
	\ddd S_s(\phi^u_s,[\phi^u]_s;u) 
		+
	\vv S_s(\phi^u_s,[\phi^u]_s;u) 
		\alpha^u(s,\phi^u_s)\right.\\
	+&\left.
	\frac1{2} \tr[
		\vv^2 S_s(\phi^u_s,[\phi^u]_s;u) 
	](\gamma^u(s,\phi_s^u))^2
\right]ds\\
+ & \int_0^t 
\vv S_s(\phi^u_s,[\phi^u]_s;u)
	\gamma^u(s,\phi_s^u)
dW_s
.
\end{aligned}
\label{eq_prop_NA_characterization_1}
\end{equation}
From the Martingale Representation Theorem, \cite[Theorem 5.2]{MartingaleRepFunctionalItoFournieRama}, and~\eqref{eq_prop_NA_characterization_1} it follows that, for each $u \in \uuu$, the process $\phi^u_t$ is a $\pp$-local-martingale if and only if 
\hspace{-.5em}
\begin{equation}
-\ddd S_s(\phi^u_s,[\phi^u]_s;u)  =
\vv S_s(\phi^u_s,[\phi^u]_s;u) 
\alpha^u(s,\phi^u_s)
+
\frac1{2} \tr[
\vv^2 S_s(\phi^u_s,[\phi^u]_s;u) 
](\gamma^u(s,\phi_s^u))^2
\label{eq_prop_NA_characterization_2}
\end{equation}

Next, the quantities $a^u$ and $\gamma^u$ are described.  By the It\^{o} formula for Riemannian manifolds with time-dependent connection, see \citep[Corollary 3.6]{guo2015martingales}, it follows that for each $u \in \uuu$ (working in local-coordinates)
\begin{equation}
\begin{aligned}
\phi(t,\beta_t,u) 
= &
\int_0^t 
\frac{\partial \phi}{\partial t}(s,\beta_s,u) 
ds
+
\int_0^t
\sum_{i=1}^d \frac{\partial \phi}{\partial\beta^i}(s,\beta_s,u)db_s^i\\
+ &
\int_0^t
\frac1{2}
\sum_{i,j=1}^{d}
\left(
\frac{\partial^2 \phi}{\partial\beta^i\partial\beta^j}
(s,\beta_s,u)
-
\sum_{k=1}^{d}\Gamma^k_{i,j}(t)
\frac{\partial^k \phi}{\partial\beta^k}
(s,\beta_s,u)
\right)
d[b]^{i,j}_s
ds
\\
= &
\int_0^t 
\left[
\frac{\partial \phi}{\partial t}(s,\beta_s,u) 
+
\sum_{i=1}^d \frac{\partial \phi}{\partial\beta^i}(s,\beta_s,u)\mu^i(s,b_s)ds
\right.
\\
+ &\left.
\frac1{2}
\sum_{i,j=1}^{d}
\left(
\frac{\partial^2 \phi}{\partial\beta^i\partial\beta^j}
(s,\beta_s,u)
-
\sum_{k=1}^{d}\Gamma^k_{i,j}(t)
\frac{\partial^k \phi}{\partial\beta^k}
(s,\beta_s,u)
\right)
\sigma^i(s,b_s)\sigma^j(s,b_s)
\right]
ds\\
+ &
\int_0^t\sum_{i=1}^d \frac{\partial \phi}{\partial\beta^i}(s,\beta_s,u)\sigma^i(s,b_s)dW^i_s
.
\end{aligned}
\label{eq_prop_NA_characterization_3}
\end{equation}
The Martingale Representation Theorem and~\eqref{eq_prop_NA_characterization_3} imply that
\begin{equation}
\begin{aligned}
\alpha^u(s,\phi^u_s)= &
\frac{\partial \phi}{\partial t}(s,\beta_s,u) 
+
\sum_{i=1}^d \frac{\partial \phi}{\partial\beta^i}(s,\beta_s,u)\mu^i(s,b_s)
\\
+ &
\frac1{2}
\sum_{i,j=1}^{d}
\left(
\frac{\partial^2 \phi}{\partial\beta^i\partial\beta^j}
(s,\beta_s,u)
-
\sum_{k=1}^{d}\Gamma^k_{i,j}(t)
\frac{\partial^k \phi}{\partial\beta^k}
(s,\beta_s,u)
\right)
\sigma^i(s,b_s)\sigma^j(s,b_s)
\\
\gamma^u(s,\phi^u_s)= &\sum_{i=1}^d \frac{\partial \phi}{\partial\beta^i}(s,\beta_s,u)\sigma^i(s,b_s)
.
\end{aligned}
\label{eq_prop_NA_characterization_4_drift_volatility_descriptions}
\end{equation}
Incorporating~\eqref{eq_prop_NA_characterization_4_drift_volatility_descriptions} into~\eqref{eq_prop_NA_characterization_2} yields~\eqref{eq_thrm_NA_condition}.  
Therefore, for $S_t(\phi^u_t,[\phi^u]_t,u)$ is a $\pp$-local-martingale, simultaneously for every $u \in \uuu$, if and only if $\pp$-a.s.~\eqref{eq_thrm_NA_condition} holds simultaneously for every $u \in \uuu$.  
\end{proof}
The drift condition obtained in Theorem~\eqref{prop_NA_characterization} implies that if $\phi$ is such that the difference of the left and right-hand sides of~\eqref{eq_thrm_NA_condition} is equal to $0$, $\pp$-a.s. for all $u \in \uuu$ then $S_t(\phi_t^u,[\phi^u]_t ;u)$ is a $\pp$-local martingale simultaneously for all $u \in \uuu$.  Thus, $S_t(\phi_t^u,[\phi^u]_t ;u)$ is simultaneously a $\pp$-local martingale for all $u \in \uuu$ if for every $u \in \uuu$ the $[0,\infty)$-valued process $\underline{\Lambda}^u_t(\phi)$ is equal to $0$ $\pp$-a.s, where $\underline{\Lambda}_t^u(\phi)$ is defined using~\eqref{eq_thrm_NA_condition} by
\begin{equation}
\begin{aligned}
\underline{\Lambda}_t^u(\phi) 
	\triangleq&
\left|
\ddd S_s(\phi^u_s,[\phi^u]_s;u)+
\vv S_s(\phi^u_s,[\phi^u]_s;u) 
\left[
\frac{\partial \phi}{\partial t}(s,\beta_s,u) 
+
\sum_{i=1}^d \frac{\partial \phi}{\partial\beta^i}(s,\beta_s,u)\mu^i(s,b_s)
\right.\right.\\
+ &
\left.\left.
\frac1{2}
\sum_{i,j=1}^{d}
\left(
\frac{\partial^2 \phi}{\partial\beta^i\partial\beta^j}
(s,\beta_s,u)
-
\sum_{k=1}^{d}\Gamma^k_{i,j}(t)
\frac{\partial^k \phi}{\partial\beta^k}
(s,\beta_s,u)
\right)
\sigma^i(s,b_s)\sigma^j(s,b_s)
\right]\right.
\\
+ &\left.
\frac1{2} \tr[
\vv^2 S_s(\phi^u_s,[\phi^u]_s;u) 
]\left(
\sum_{i=1}^d \frac{\partial \phi}{\partial\beta^i}(s,\beta_s,u)\sigma^i(s,b_s)
\right)^2
\right|
;
\end{aligned}
\label{eq_definition_AF_2_description_left_m_right}
\end{equation}
where, as before, the abbreviation $\phi^u_t\triangleq  
\phi(t,\beta_t,u)$ was made.  

The processes $\underline{\Lambda}^u_t(\phi)$ can be used to define the penalty in~\eqref{AFReg}, by integrating overall values of $t$ and $u$.  However, in certain applications, such as with bond-markets, it is more convenient to instead build $\AF^{\lambda}$ from a divisor of $\underline{\Lambda}^u_t(\phi)$ as formalized by the following definition.  
\begin{defn}[Arbitrage-Penalty]\label{defn_AF_Pen}
Let $\{\Lambda^u_t(\phi)\}_{\phi \in \hhh;u \in \uuu}$ be a family of $\fff_t$-adapted $[0,\infty)$-valued stochastic processes for which
\begin{equation}
\begin{aligned}
	\Lambda^u_t(\phi)(\omega) =0 
&\Rightarrow
	\underline{\Lambda}^u_t(\phi)(\omega)=0 
%
,
\end{aligned}
\label{defn_AF_Pen_Pas_zero_sets}
\end{equation}
holds for all $\phi \in \hhh$, $t \in I$, $u \in \uuu$, and $\pp$-almost every $\omega \in \Omega$.  Then, for every $\lambda\geq0$, the family $\{\AF^{\lambda}\}_{\lambda \geq 0}$ of functions
\begin{equation}
\begin{aligned}
&\AF^{\lambda}: \, \xxx  \rightarrow [0,\infty]
\\
&\phi  \mapsto
\lambda
\ee{
	\sqrt[\lambda]{\int_{(t,u) \in I\times \uuu} |\Lambda^u_t(\phi)|^{\lambda} d\mu(t,u)}
}
.
\end{aligned}
\label{defn_AF_Penalty_after_integration_lambda_process}
\end{equation}
is said to define an \textit{arbitrage-penalty}.  
\end{defn}
\begin{remark}
The convention that the integral $\ee{
	\sqrt[\lambda]{\int_{(t,u) \in I\times \uuu} |\Lambda^u_t(\phi)|^{\lambda} d\mu(t,u)}
}$ equals to (positive) $\infty$ whenever its integrand fails to be integrable, will be maintained throughout this paper.  
\end{remark}
The equivalence between problems~\eqref{AF} and~\eqref{AFReg} is demonstrated in the next theorem.  The proof relies on the theory of $\Gamma$-convergence.  This theory is devoted to the analysis of problems requiring the interchanging of a limit and an $\operatorname{arginf}$ operation.  
Briefly, the theory of $\Gamma$-convergence is an extrapolation of Weirestrass's Theorem, \cite[Theorem 2.2]{focardi2012gamma}.  This fundamental result of optimization theory, states that a proper, lower semi-continuous, and coercive function that is bounded-below admits a minimizer.  
Extrapolating, it is shown in \cite{de1980gamma}, that if a sequence of proper functions converges to a proper lower semi-continuous function which is coercive in an appropriately uniform sense, called \textit{equicoercivity}, then the sequence of minimizers of the sequence of function, converges to a minimizer of the limiting function.  

Alternatively, the $\Gamma$-convergence of a sequence of functions can be understood geometrically, as the Kuratowski convergence (a form of set-convergence introduced in \cite{KuratowskiTopology}) of the function sequence's epigraphs to the epigraph of the limiting function.  The convergence of the epigraphs then implies that the lowest point of each of those epigraphs converges to the lowest-point of the closure of the limiting function's epigraph.  Moreover, any such lowest-point must lie on the limiting function's epigraph if it is lower semi-continuous.  Thus, in this case, the limit and $\operatorname{arginf}$ operations can be interchanged.  Further details on $\Gamma$-convergence can also be found in \cite{dal2012introduction,kratsios2019partial}.  The following additional assumption, which is satisfied in the term-structure setting, is required.  
\begin{ass}\label{ass_for_Gamma_Convergence}
  The following assumptions will be required.  \hfill
\begin{enumerate}[(i)]
\item For every $\phi \in \hhh$ and $\pp$-a.e. $\omega\in \Omega$, the function $(t,u)\mapsto \Lambda^u_t(\phi)(\omega)$ is continuous on $\hhh$,
\item $\left\{
\phi \in \hhh
		:\, 
	(\forall u\in \uuu) 
		\, 
		S_t(\phi^u_{t},[\phi^u]_{t};u) \, \mbox{is a $\pp$-local-martingale}
\right\}\subseteq \hhh$ is closed and non-empty.
\end{enumerate}
Note that both statements (i) and (ii) are with respect to the relative topology on $\hhh$.  
\end{ass}
\begin{thrm}\label{thrm_Gamma_limit}
  Under Assumption~\ref{ass_for_Gamma_Convergence} the following hold: 
\begin{enumerate}[(i)]
\item \eqref{AF} admits a minimizer on $\hhh$
\item $\lim\limits_{\lambda \uparrow \infty;\, \lambda\geq 2}
\inf_{\phi \in \hhh}
\ell(\varphi - \phi)
+
\AF^{\lambda}(\phi)
=
\min_{\phi \in \hhh}
\ell(\varphi-\phi)
+
\iota_{\hhh}(\phi)
,
$
\item If for every $\lambda\geq 2$ $\AF^{\lambda}$ is lower-semi-continuous on $\hhh$ then
\begin{equation}
\lim\limits_{\lambda \uparrow \infty;\, \lambda\geq 2}
\underset{\phi \in \hhh}{\operatorname{argmin}}
		\,%
\ell(\varphi) 
+
\AF^{\lambda}(\phi)
=
\underset{\phi \in \hhh}{\operatorname{argmin}}
	\,%
\ell(\varphi) 
+
\iota_{\hhh}(\phi)
.
\label{thrm_Gamma_limit_statement_partb_convergence_of_minimizers}
\end{equation}
Where $\iota_{\hhh}$ is defined on $\hhh$ as
\begin{equation}
\begin{aligned}
\iota_{\hhh}(\phi)&\triangleq 
\begin{cases}
0 &: (\forall u \in \uuu)\, S_t(\phi^u_t,[\phi^u]_t;u) \mbox{ is a $\pp$-local-martingale}
\\
\infty &:\, \mbox{else}
.
\end{cases}
\end{aligned}
\label{thrm_Gamma_limit_definition_of_Plocal_martingale_detector_indicator_function}
\end{equation}
\end{enumerate}
\end{thrm}
\begin{proof}
Since $(I\times \uuu,\mathcal{B}\left(I\times \uuu\right),\mu)$ is a finite measure space then
	\begin{equation}
\lim\limits_{\lambda \uparrow \infty; \, \lambda\geq 2}
\left(
	\int_{(t,u) \in I\times \uuu} |\Lambda^u_t(\phi)|^{\lambda} d\mu(t,u)
\right)^{\frac1{\lambda}}
=
\underset{
	{(t,u)\in I\times \uuu}
}{
	\operatorname{esssup}
}
\Lambda^u_t(\phi)
;
\label{thrm_Gamma_limit_eq_1_pw_convergence}
\end{equation}
By Assumption~\eqref{ass_for_Gamma_Convergence} (i), the map $(t,u)\mapsto \Lambda^u_t(\phi)$ is continuous, for each $\phi \in \hhh$, therefore
\begin{equation}
\underset{
	{(t,u)\in I\times \uuu}
}{
	\operatorname{esssup}
}
\Lambda^u_t(\phi)
	=
\underset{
	{(t,u)\in I\times \uuu}
}{
	\operatorname{sup}
}
\Lambda^u_t(\phi)
\label{thrm_Gamma_limit_eq_3_esssup_is_sup}
.
\end{equation}
Since the product of limits is the limit of the product, then~\eqref{thrm_Gamma_limit_eq_1_pw_convergence} yields
\begin{equation}
\begin{aligned}
\lim\limits_{\lambda \uparrow \infty; \, \lambda\geq 2} 
\lambda
\left(
	\int_{(t,u) \in I\times \uuu} |\Lambda^u_t(\phi)|^{\lambda} d\mu(t,u)
\right)^{\frac1{\lambda}}
= &
\lim\limits_{\lambda \uparrow \infty; \, \lambda\geq 2} 
\lambda
\underset{
	{(t,u)\in I\times \uuu}
}{
	\operatorname{esssup}
}
\Lambda^u_t(\phi)
\\
=&
\lim\limits_{\lambda \uparrow \infty; \, \lambda\geq 2} 
\lambda
\underset{
	{(t,u)\in I\times \uuu}
}{
	\operatorname{sup}
}
\Lambda^u_t(\phi)\\
= & 
\begin{cases}
0 &: \, 	
\underset{
	{(t,u)\in I\times \uuu}
}{
	\operatorname{sup}
}
\Lambda^u_t(\phi)
	=0
	\\
\infty &:\, \mbox{else}
.
\end{cases}
\end{aligned}
\label{thrm_Gamma_limit_eq_4_monotone_convergence}
\end{equation}
Since $\mu$ is a probability measure in $I\times \uuu$ then for every $1\leq \lambda_1\leq \lambda_2< \infty$, it follows that for every $\phi \in \hhh$ 
\begin{equation}
\left(
	\int_{(t,u) \in I\times \uuu} |\Lambda^u_t(\phi)|^{\lambda_1} d\mu(t,u)
	\right)^{\frac1{\lambda_1}}
\leq
\left(\int_{(t,u) \in I\times \uuu} |\Lambda^u_t(\phi)|^{\lambda_2} d\mu(t,u)
\right)^{\frac1{\lambda_2}}
\leq
\underset{
	{(t,u)\in I\times \uuu}
}{
	\operatorname{esssup}
}
\left|
\Lambda^u_t(\phi)
\right|
.
\label{thrm_Gamma_limit_eq_2_monotonicity_pt1}
\end{equation}
Thus, for every $\phi \in \hhh$, the convergence described by~\eqref{thrm_Gamma_limit_eq_1_pw_convergence} is monotone (increasing) and non-negative; therefore by the monotone convergence theorem, it follows that
\begin{equation}
\begin{aligned}
\lim\limits_{\lambda \uparrow \infty; \, \lambda\geq 2} 
\lambda\ee{
\left(
	\int_{(t,u) \in I\times \uuu} |\Lambda^u_t(\phi)|^{\lambda} d\mu(t,u)
\right)^{\frac1{\lambda}}
}
= &
\begin{cases}
0 &: \, 	
\underset{
	{(t,u)\in I\times \uuu}
}{
	\operatorname{sup}
}
\Lambda^u_t(\phi)
=0, \, \pp-a.s.
\\
\infty &:\, \mbox{else}
.
\end{cases}
\end{aligned}
\label{thrm_Gamma_limit_eq_4a_monotone_convergence}
\end{equation}

Applying Proposition~\ref{prop_NA_characterization} to the right-hand side of~\eqref{thrm_Gamma_limit_eq_4a_monotone_convergence}, it follows that
\begin{equation} 
\iota_{\hhh}(\phi)
	=
\begin{cases}
0 &: \, 	
\underset{
	{(t,u)\in I\times \uuu}
}{
	\operatorname{sup}
}
\Lambda^u_t(\phi), \, \pp-a.s.
=0
\\
\infty &:\, \mbox{else}
\end{cases}
\label{thrm_Gamma_limit_eq_5_pw_limit_to_Plocalmartingale_detector}
\end{equation}
where $\iota_{\hhh}$ is defined as in~\eqref{thrm_Gamma_limit_definition_of_Plocal_martingale_detector_indicator_function}.  
Therefore, the following limit holds
\begin{equation}
\begin{aligned}
\lim\limits_{\lambda \uparrow \infty; \, \lambda\geq 2} 
\lambda
\ee{
	\left(
		\int_{(t,u) \in I\times \uuu} |\Lambda^u_t(\phi)|^{\lambda} d\mu(t,u)
	\right)^{\frac1{\lambda}}
}
	&= 
\iota_{\hhh}(\phi)
&\qquad
\left(
\forall \phi \in \hhh
\right)
.
\end{aligned}
\label{thrm_Gamma_limit_eq_6_pw_convergence_completed}
\end{equation}
Thus~\eqref{thrm_Gamma_limit_eq_6_pw_convergence_completed} establishes the convergence of the penalty functions $\AF^{\lambda}$ to $\iota_{\hhh}$, in $\hhh$.  Next, their $\Gamma$-convergence is established and their $\Gamma$-convergence is used to deduce the $\Gamma$-convergence of the objective functions in~\eqref{AFReg} to the objective function of~\eqref{AF}.

Applying~\eqref{thrm_Gamma_limit_eq_2_monotonicity_pt1} and the monotonicity of integration, it follows that for every $\phi \in \hhh$
\begin{equation}
\hspace*{-1em}
\lambda_1\ee{
	\left(
		\int_{(t,u) \in I\times \uuu} |\Lambda^u_t(\phi)|^{\lambda_1} d\mu(t,u)
	\right)^{\frac1{\lambda_1}}
}
\leq
\lambda_2\ee{
	\left(\int_{(t,u) \in I\times \uuu} |\Lambda^u_t(\phi)|^{\lambda_2} d\mu(t,u)
	\right)^{\frac1{\lambda_2}}
}
\leq 
\iota_{\hhh}(\phi)
.
\label{thrm_Gamma_limit_eq_2_monotonicity_pt1_alt}
\end{equation}
Thus,~\eqref{thrm_Gamma_limit_eq_2_monotonicity_pt1_alt} together with \citep[Proposition 5.4]{dal2012introduction} and~\citep[Remark 1.40 (ii)]{braides2002gamma} imply that (on $\hhh$)
\begin{equation}
\GamLim{\lambda \uparrow \infty; \, \lambda\geq 2} 
\lambda
\ee{
	\left(\int_{(t,u) \in I\times \uuu} |\Lambda^u_t(\phi)|^{\lambda} d\mu(t,u)
	\right)^{\frac1{\lambda}}
}
= 
\iota_{\hhh}^{lsc}(\phi)
,
\label{thrm_Gamma_limit_eq_7_Gamma_convergence_pt1_lsc}
\end{equation}
where $\iota_{\hhh}^{lsc}$ is the lower-semi-continuous relaxation of $\iota_{\hhh}$ on $\hhh$; that is, the smallest lower-semi-continuous function dominating $\iota_{\hhh}$ on $\hhh$; a precise description can be found on \citep[page 11]{focardi2012gamma}.  However, Assumption~\eqref{ass_for_Gamma_Convergence} (ii) implies that $\iota_{\hhh}$ is indeed lower-semi-continuous; thus $\iota_{\hhh}^{lsc}=\iota_{\hhh}$, on $\hhh$.  Therefore~\eqref{thrm_Gamma_limit_eq_7_Gamma_convergence_pt1_lsc} simplifies (on $\hhh$) to
\begin{equation}
\GamLim{\lambda \uparrow \infty; \, \lambda\geq 2} 
\lambda
\ee{
	\left(
		\int_{(t,u) \in I\times \uuu} |\Lambda^u_t(\phi)|^{\lambda} d\mu(t,u)
	\right)^{\frac1{\lambda}}
}
= 
\iota_{\hhh}(\phi)
.
\label{thrm_Gamma_limit_eq_8_Gamma_convergence_pt2_no_lsc}
\end{equation}
Since $\Gamma$-limits are invariant under continuous perturbation, see \citep[Theorem 2.8.]{focardi2012gamma}, then~\eqref{thrm_Gamma_limit_eq_8_Gamma_convergence_pt2_no_lsc} and the continuity of $\ell(\varphi- \cdot)$ on $\hhh$ implies that (on $\hhh$)
\begin{equation}
\GamLim{\lambda \uparrow \infty; \, \lambda\geq 2} 
\ell\left(\varphi - \cdot\right)
+
\lambda
\ee{
	\left(\int_{(t,u) \in I\times \uuu} |\Lambda^u_t(\cdot)|^{\lambda} d\mu(t,u)
	\right)^{\frac1{\lambda}}
}
= 
\ell(\varphi) 
+
\iota_{\hhh}(\cdot)
.
\label{thrm_Gamma_limit_eq_9_Gamma_convergence_pt3_objective_functions}
\end{equation}

In order to apply the Fundamental Theorem of $\Gamma$-convergence, the family of functions on the left-hand side of~\eqref{thrm_Gamma_limit_eq_9_Gamma_convergence_pt3_objective_functions} must be equicoercive.  Since $\lambda
\ee{
	\left(\int_{(t,u) \in I\times \uuu} |\Lambda^u_t(\cdot)|^{\lambda} d\mu(t,u)
	\right)^{\frac1{\lambda}}
}$ is non-negative and since $\hhh$ is unbounded then $\ell(\varphi) 
$ is coercive on $\hhh$, ie:
\begin{equation}
\lim\limits_{\lambda\uparrow \infty;\,\lambda\geq 2}
	\ell(\varphi) 
	 = \infty
\label{thrm_Gamma_limit_eq_10_equicoercive_coerciveness_norm_for_lb}
,
\end{equation}
and by \citep[Proposition 7.7]{dal2012introduction} together with~\eqref{thrm_Gamma_limit_eq_10_equicoercive_coerciveness_norm_for_lb} it follows that
\begin{equation}
\begin{aligned}
\ell(\varphi) 
&\leq 
\ell(\varphi) 
+
\lambda
\ee{
	\left(\int_{(t,u) \in I\times \uuu} |\Lambda^u_t(\cdot)|^{\lambda} d\mu(t,u)
\right)^{\frac1{\lambda}}
}
&
\qquad \left(
\forall \lambda \geq 2
\right)
;
\end{aligned}
\label{thrm_Gamma_limit_eq_11_equicoercive_using_coercive_lb}
\end{equation}
whence, $\left\{
	\ell(\varphi) 
	+
	\lambda
	\ee{
		\left(\int_{(t,u) \in I\times \uuu} |\Lambda^u_t(\cdot)|^{\lambda} d\mu(t,u)
		\right)^{\frac1{\lambda}}
	}
\right\}_{\lambda \geq 2}$ forms an equicoercive family, on $\hhh$.  

Thus, $\left\{
\ell(\varphi) 
+
\lambda
\ee{
	\left(\int_{(t,u) \in I\times \uuu} |\Lambda^u_t(\cdot)|^{\lambda} d\mu(t,u)
	\right)^{\frac1{\lambda}}
}
\right\}_{\lambda \geq 2}$ defines an equicoercive family which $\Gamma$-converges to $\iota_{\hhh}$, on $\hhh$.  Therefore, together~\eqref{thrm_Gamma_limit_eq_9_Gamma_convergence_pt3_objective_functions} and~\eqref{thrm_Gamma_limit_eq_11_equicoercive_using_coercive_lb} imply that the Fundamental Theorem of $\Gamma$-convergence, \citep[Theorem 7.8]{dal2012introduction}
, applies.  Hence, 
\begin{equation}
\lim\limits_{\lambda \uparrow \infty;\, \lambda\geq 2}
\inf_{\phi \in \hhh}
\ell(\varphi) 
+
\lambda
\ee{
	\left(\int_{(t,u) \in I\times \uuu} |\Lambda^u_t(\phi)|^{\lambda} d\mu(t,u)
	\right)^{\frac1{\lambda}}
}
=
\min_{\phi \in \hhh}
\ell(\varphi) 
	+
\iota_{\hhh}(\phi)
\label{thrm_Gamma_limit_eq_12_application_of_fundamental_theorem_of_Gamma_convergence}
.
\end{equation}

Lastly, \citep[Theorem 7.8]{dal2012introduction} also implies that $\ell(\varphi) 
+
\iota_{\hhh}(\cdot)$ is coercive on $\hhh$.  Hence, $\ell(\varphi) 
+
\iota_{\hhh}(\cdot)$ is coercive, lower-semi-continuous, and bounded-below by $0$.  Therefore, by Weirestrass's Theorem, \cite[Theorem 2.2]{focardi2012gamma}, it follows that $\ell(\varphi) 
+
\iota_{\hhh}(\cdot)$ admits a minimizer on $\hhh$.  
\end{proof}
%
Next, Theorem~\ref{thrm_Gamma_limit} and the arbitrage-free regularization~\eqref{AFReg} will be applied to the bond market.  
\section{Arbitrage-Free Regularization for Bond Pricing}\label{s_AF_Reg_FRC}
As discussed in \cite{YieldDbold}, affine term-structure models are commonly used in forward-rate curve modelling due to their tractability and the interpretability.  In the formulation of \cite{bjork1999interest}, as further developed in \cite{filipovic2000exponential,FilTapTeich}, affine term-structure models are characterized by~\eqref{eq_ATS} together with the additional requirement that its stochastic factor process $\beta_t$ follows an affine diffusion; which, by results such as \cite{DuffieAffine,cuchiero2011affine,keller2008affine}, implies that the dynamics of $\beta_t$ are given by
\begin{equation}
\begin{aligned}
\mu^i(t,\beta)
	&\triangleq 
\gamma^i + \sum_{i=1}^{d} \gamma_{i,j} \beta^i
\\
\left[\sigma^i(t,\beta)\right]^T\sigma^j(t,\beta)
	& \triangleq
\alpha^{i,j} + \sum_{i,j=1}^{d}\alpha_{k;i,j}\beta^k
,
\end{aligned}
\label{eq_ATS_dynamics}
\end{equation}
where $\gamma^i,\gamma_{i,j},\alpha^{i,j},\alpha_{k;i,j}\in \rr$ and $i,j=1,\dots,d$.  Note further that, in this setting, the geometry on $\mmm=\rrd$ is Euclidean; whence the Christoffel symbols vanish, ie: $\Gamma^{k}_{i,j}=0$.  

Fix meta-parameters $p,\kappa\geq 1$.  For the next result, all the factor-models will be taken as belonging to the weighted Sobolev space $W^{p,k}_{w}(I\times \mmm\times \uuu)$ with weight function
\begin{equation}
\begin{aligned}
w(t,\beta,u)\triangleq C e^{-|t| - \|\beta\|^{\kappa}-|u|^{\kappa}} 
,
\end{aligned}
\label{eq_growth_conditions}
\end{equation}
where $C$ is a unique constants ensuring that $1\in W_w^{p,k}$ and its weighted integral is equal to $1$, and $k$ is required to satisfy
\begin{equation}
\begin{aligned}
k&\geq \frac{1+d+D}{p}+2
.
\end{aligned}
\label{eq_Regularity_Assumption}
\end{equation}

Analytic tractability is ensured by requiring that the factor-models considered, for the arbitrage-free regularization~\eqref{AFReg}, belong to the class $\hhh$ defined by
\begin{equation}
\phi(t,\beta,u)= \phi_0(u+t) +\sum_{i=1}^{d} \beta^i\phi_i(u+t)
\label{eq_class_H_FRC_regularization}
.
\end{equation}
Under these conditions, the following theorem characterizes the asymptotic behavior of~\eqref{AFReg} in $\lambda$ as solving~\eqref{AF}; given fixed meta-parameters $p,\kappa\geq 1$.  Following~\cite{filipovic2001consistency}, it will be convenient to denote
\begin{equation}
\Phi^i(u)=\int_0^t \phi^i(s)ds
\label{eq_integral_definition_big_Phi_from_little_phi}
.
\end{equation}
\begin{thrm}\label{thrm_AF_Pen_ATS}
Let $\varphi$ be in $\hhh$ and fix $p,\kappa\geq 1$.  Then 
\begin{enumerate}[(i)]
\item For every $\lambda \geq 2$ there exists an element $\phi^{\lambda}$ in $\hhh$ minimizing
$$
\int_{0}^{\infty}\int_{\beta \in \rrd} e^{-|u|^{\kappa}-\|\beta\|^{\kappa}}
\left( 
\varphi(u,\beta)
-
\phi(u,\beta)
\right)^p d\beta du
+
\frac{\lambda}{\Gamma(1+\frac1{\kappa})^{\frac1{\lambda}}}
\sqrt[\lambda]{
	\int_{0}^{\infty}
	e^{-|u|^{\kappa}}
	\left|
	\Lambda^u(\phi)
	\right|^{\lambda}
	du
}
;
$$
where $\Lambda^u_t(\phi)$ is defined by
\begin{equation}
\begin{aligned}
\Lambda_t^u(\phi)\triangleq &\left|
c_0-\frac{\partial \Phi^0}{\partial u}(u) + \sum_{i=1}^d \gamma^i\Phi^i(u) - \frac1{2}\sum_{i,j=1}^d\alpha^{i,j}\Phi^i(u)^{\star}\Phi^j(u)
\right|^p\\
+\sum_{k=1}^d 
&\left|
c_k-\frac{\partial \Phi^k}{\partial u}(u) + \sum_{i=1}^d \gamma^{k,i}\Phi^i(u) - \frac1{2}\sum_{i,j=1}^d\alpha_{k;i,j} \Phi^i(u)^{\star}\Phi^j(u)
\right|^p
.
\end{aligned}
\label{thrm_AF_Pen_ATS_definition_AF_penalty}
\end{equation}
\item The following inclusion holds
\begin{equation}
\begin{aligned}
\hspace*{-1.5em}
\lim_{\lambda \uparrow \infty;\,\lambda \geq 2}
\phi^{\lambda} &
\in
\underset{\phi \in \hhh}{\argmin}&
\int_{0}^{\infty}\int_{\beta \in \rrd} 
	e^{-|u|^{\kappa}-\|\beta\|^{\kappa}}
\left( 
\varphi(u,\beta)
-
\phi(u,\beta)
\right)^p d\beta du
	+
\iota_{\hhh}(\phi)
;
\end{aligned}
\label{thrm_AF_Pen_ATS_limittin_statement}
\end{equation}
where $\iota_{\hhh}$ is as in~\eqref{thrm_Gamma_limit_definition_of_Plocal_martingale_detector_indicator_function}. 
\end{enumerate}
\end{thrm}
\begin{proof}
By definition of $(\mmm,g_t)$, $\beta_t$, and $\hhh$ Assumption~\ref{assumptions_summary} (i), (iv), and (v) hold; thus only Assumption~\ref{assumptions_summary} (ii) and (iii) must be verified, in order to ensure that the stated problem falls within the scope of this paper.  Let $\mu=\tilde{\mu}\otimes \upsilon$ where $\frac{d\tilde{\mu}}{dm}(u)=
\frac{e^{-|u|^{\kappa}}}{\Gamma\left(1+\frac1{\kappa}\right)}I_{[0,\infty)}
$, $\upsilon$ is the unique probability measure with Lebesgue density proportional to $e^{-\|\beta\|^{\kappa}}$ on $\rrd$, and where $\frac{d \nu}{dm}(t)=e^{-|t|}1_{[0,\infty)}$, $1_{[0,\infty)}$ is the (probabilistic not convex analytic) indicator function on the interval $[0,\infty)$, here $m$ is the Lebesgue measure on $\rr$.  Therefore, the elements of $W_w^{p,k}(I \times \mmm\times \uuu)$ are elements of $L_{\nu \otimes \mu}^p(I\times \mmm\times \uuu)$.  Since $[0,\infty)\times [0,\infty)\times \rrd$ has a smooth boundary and since $k$ was assumed to satisfy~\eqref{eq_Regularity_Assumption}, then the (weighted) Morrey-Sobolev Theorem of \cite{SobolevEmbeddingWeightedOpicBrown}  applies.  Therefore, $W^{p,k}_{w}(I\times \mmm\times \uuu)$ can be continuously embedded within $C^{2}(I\times \mmm\times \uuu)$ and therefore Assumption~\ref{assumptions_summary} (ii) holds.  Furthermore, by~\eqref{eq_structure_map_bond_case} and together with \citep[Example 1]{MartingaleRepFunctionalItoFournieRama} each $S_t(\cdot,\cdot;u)$ satisfies Assumption~\ref{assumptions_summary} (iii); thus Assumption~\ref{assumptions_summary} is satisfied.

Next,~\eqref{thrm_AF_Pen_ATS_limittin_statement} is reformulated in terms of Theorem~\eqref{thrm_Gamma_limit} and Assumptions~\ref{ass_for_Gamma_Convergence} are verified.  Subsequently, the optimizers of the objective-function under the limit on the left-hand side of~\eqref{thrm_AF_Pen_ATS_limittin_statement} are shown to exist for $\lambda\geq 2$.

In the case where each $S_t(\cdot,\cdot;u)$ is as in~\eqref{eq_structure_map_bond_case}, it is shown in \citep[]{filipovic2001consistency} that for each $u \in \uuu$ the bond prices $S_t\left(
\phi_t^u,[\phi^u]_t ;u
\right)$ are each $\pp$-local martingales if and only if for every $u \in \uuu$, $t \in I$, and $\pp$-a.e $\omega \in \Omega$, the following holds
\begin{equation}
\begin{aligned}
0=&\left(
\phi_0(0)-\phi_0(u) + \sum_{i=1}^d \gamma^i\Phi^i(u) - \frac1{2}\sum_{i,j=1}^d\alpha^{i,j}\Phi^i(u)^{\star}\Phi^j(u)
\right)\\
+\sum_{k=1}^d \beta_t^k
&
\left(
\phi_k(0)-\phi_k(u) + \sum_{i=1}^d \gamma^{k,i}\Phi^i(u) - \frac1{2}\sum_{i,j=1}^d\alpha_{k;i,j} \Phi^i(u)^{\star}\Phi^j(u)
\right)
\end{aligned}
\label{proof_prop_AF_Pen_ATS_2_HJM_Drift_Ito_Simplication_of_consistency_condition}
\end{equation}
Equation~\eqref{proof_prop_AF_Pen_ATS_2_HJM_Drift_Ito_Simplication_of_consistency_condition} is satisfied for $\pp$-a.s every $\omega \in \Omega$, for every $t \in I$, and for every $u \in \uuu$ if the family of deterministic "processes" $\{\Lambda_t^u(\phi)\}_{u \in \uuu}$ are $\pp$-a.s. $0$; where $\Lambda_t^u(\phi)$ is defined by
\begin{equation}
\begin{aligned}
\Lambda_t^u(\phi)\triangleq &\left|
c_0-\frac{\partial \Phi^0}{\partial u}(u) + \sum_{i=1}^d \gamma^i\Phi^i(u) - \frac1{2}\sum_{i,j=1}^d\alpha^{i,j}\Phi^i(u)^{\star}\Phi^j(u)
\right|^p\\
+\sum_{k=1}^d 
&\left|
c_k-\frac{\partial \Phi^k}{\partial u}(u) + \sum_{i=1}^d \gamma^{k,i}\Phi^i(u) - \frac1{2}\sum_{i,j=1}^d\alpha_{k;i,j} \Phi^i(u)^{\star}\Phi^j(u)
\right|^p
.
\end{aligned}
\label{proof_prop_AF_Pen_ATS_5_reduction_to_sufficient_condition}
\end{equation}
Therefore, since $\Lambda_t^u(\phi)$ satisfies~\eqref{defn_AF_Pen_Pas_zero_sets} then the family $\{AF^{\lambda}\}_{\lambda>0}$ of functions defined by
\begin{equation}
AF^{\lambda}(\phi)\triangleq 
\lambda
\ee{
	\sqrt[\lambda]{
		\int_{0}^{\infty}
		\int_{\beta \in \rrd}
			\int_0^{\infty}
			\frac1{\Gamma(1+\frac1{\kappa})}
				e^{-|u|^{\kappa}}
					\left|
						\Lambda^u_t(\phi)
					\right|^{\lambda}
				d\tilde{\mu}
				d\upsilon(\beta)
			d\nu{s}
	}
}
,
\label{proof_prop_AF_Pen_ATS_6_arbitrage_penalty_construction}
\end{equation}
define an arbitrage-penalty in the sense of~\eqref{defn_AF_Penalty_after_integration_lambda_process}.  Since, $\Lambda_t^u(\phi)$ is $\pp$-a.s. deterministic and constant both in $t$ and in $\beta$, then~\eqref{proof_prop_AF_Pen_ATS_6_arbitrage_penalty_construction} further simplifies to
\begin{equation}
AF^{\lambda}(\phi)= 
\frac{\lambda}{\Gamma(1+\frac1{\kappa})^{\frac1{\lambda}}}
	\sqrt[\lambda]{
		\int_{0}^{\infty}
		e^{-|u|^{\kappa}}
		\left|
		\Lambda^u_t(\phi)
		\right|^{\lambda}
		du
	}
\label{proof_prop_AF_Pen_ATS_7_arbitrage_penalty_simplified_form_removed_integrals}
\end{equation}
Since $W_w^{k,p}(I\times \mmm\times \uuu)$ is continuously embedded in $C^2(I\times \mmm\times \uuu)$, then each equivalence class $\phi \in W_w^{k,p}(I\times \mmm\times \uuu)$ can be identified with a continuous function from $I\times \mmm\times \uuu$; therefore each of the functions
\begin{equation}
\begin{aligned}
u\mapsto &\left|
c_0-\frac{\partial \Phi^0}{\partial u}(u) + \sum_{i=1}^d \gamma^i\Phi^i(u) - \frac1{2}\sum_{i,j=1}^d\alpha^{i,j}\Phi^i(u)^{\star}\Phi^j(u)
\right|^p\\
u\mapsto &\left|
c_k-\frac{\partial \Phi^k}{\partial u}(u) + \sum_{i=1}^d \gamma^{k,i}\Phi^i(u) - \frac1{2}\sum_{i,j=1}^d\alpha_{k;i,j} \Phi^i(u)^{\star}\Phi^j(u)
\right|^p
,
\end{aligned}
\label{proof_prop_AF_Pen_ATS_8_continuityinuandt}
\end{equation}
are continuous in $u$; moreover, they are continuous in $t$ since they are constant in $t$.  Therefore, $(t,u)\mapsto \Lambda_t^u(\phi)$ is continuous for every $\phi \in \hhh$; whence Assumption~\ref{ass_for_Gamma_Convergence} (i) holds.

Next, Assumption~\ref{ass_for_Gamma_Convergence} (ii) will be verified.  Given the dynamics of~\eqref{eq_ATS_dynamics}, \citep[Proposition 9.3]{filipovicTSMbookypoop} characterizes all $\phi_0,\dots,\phi_d
$ for which the forward-rate curve~\eqref{eq_class_H_FRC_regularization} corresponds to a bond market, through~\eqref{eq_structure_map_bond_case}, in which each bond price is a $\pp$-local-martingale; all such $\phi_0,\dots,\phi_d$ are solutions to the differential Riccati system
\begin{equation}
\begin{aligned}
\frac{\partial \Phi^0}{\partial u}(u) &= c_0 + \sum_{i=1}^d \gamma^i\Phi^i(u) - \frac1{2}\sum_{i,j=1}^d\alpha^{i,j}\Phi^i(u)^{\star}\Phi^j(u)
&\qquad  \Phi^0(0)=0
&\\
\frac{\partial \Phi^k}{\partial u}(u)&=
c_k + \sum_{i=1}^d \gamma^{k,i}\Phi^i(u) - \frac1{2}\sum_{i,j=1}^d\alpha_{k;i,j} \Phi^i(u)^{\star}\Phi^j(u)
&\qquad  \Phi^k(0)=0
&;
\end{aligned}
\label{proof_prop_AF_Pen_ATS_8a_descritption_of_solutions_in_terms_of_riccatti_system}
\end{equation}
where $c_0,\dots,c_k$ are any elements of $\rr$.  Thus, 
\begin{equation}
\begin{aligned}
&\left\{
\phi \in \hhh
:\, 
(\forall u\in \uuu) 
\, 
S_{t}(\phi^u_{t},[\phi^u]_{t};u) \, \mbox{is a $\pp$-local-martingale}
\right\}
\\
=&
\left\{
\phi \in \hhh
:\, 
(\exists c_0,\dots,c_d \in \rr) 
\, 
\{\Phi_i\}_{i=0}^d \mbox{{ solves~\eqref{proof_prop_AF_Pen_ATS_8a_descritption_of_solutions_in_terms_of_riccatti_system}}}
\right\}
,
\end{aligned}
\label{proof_prop_AF_Pen_ATS_8b_equivalentce_between_Riccatti_and_local_martingale}
\end{equation}
where as before, $\{\Phi^i\}_{i=0}^d$ and $\phi$ are related through~\eqref{eq_class_H_FRC_regularization} and~\eqref{eq_integral_definition_big_Phi_from_little_phi}.  Differentiating across the Riccatti system~\eqref{proof_prop_AF_Pen_ATS_8a_descritption_of_solutions_in_terms_of_riccatti_system} with respect to $u$ yields an equivalent differential system of the form
\begin{equation}
\begin{aligned}
&\frac{\partial^2 \Phi^0}{\partial u^2}(u) = &c_0 + \sum_{i=1}^d \gamma^i\frac{\partial \Phi^i}{\partial u} 
&- \frac1{2}\sum_{i,j=1}^d\alpha^{i,j}
\left[\frac{\partial \Phi^i}{\partial u}\Phi^j(u) + \Phi^i(u)\frac{\partial \Phi^j}{\partial u}\right],
&\\
&\frac{\partial^2 \Phi^k}{\partial u^2}(u)=
&c_k + \sum_{i=1}^d \gamma^{k,i}\frac{\partial \Phi^i}{\partial u} 
&- \frac1{2}\sum_{i,j=1}^d\alpha_{k;i,j} \left[\frac{\partial \Phi^i}{\partial u}\Phi^j(u) + \Phi^i(u)\frac{\partial \Phi^j}{\partial u}\right],\\
&  \Phi^0(0)=0,
&\,  \Phi^k(0)=0,\\
& \frac{\partial \Phi^0}{\partial u}(0)=c_0 
& \, \frac{\partial \Phi^k}{\partial u}(0)=c_k;\\
\end{aligned}
\label{proof_prop_AF_Pen_ATS_8c_Riccatti_second_order_linearized}
\end{equation}
Therefore,~\eqref{proof_prop_AF_Pen_ATS_8b_equivalentce_between_Riccatti_and_local_martingale} can be rewritten as
\begin{equation}
\begin{aligned}
&\left\{
\phi \in \hhh
:\, 
(\forall u\in \uuu) 
\, 
S_{t}(\phi^u_{t},[\phi^u]_{t};u) \, \mbox{is a $\pp$-local-martingale}
\right\}
\\
=&
\left\{
\phi \in \hhh
:\, 
(\exists c_0,\dots,c_d \in \rr) 
\, 
\{\Phi_i\}_{i=0}^d \mbox{{ solves~\eqref{proof_prop_AF_Pen_ATS_8c_Riccatti_second_order_linearized}}}
\right\}
.
\end{aligned}
\label{proof_prop_AF_Pen_ATS_8d_equivalentce_between_Riccatti_and_local_martingale_constrained_Version}
\end{equation}
Since the $C^k(\rr;\rr^{d+1})$ is equipped with its the topology of uniform convergence on compacts of functions and their first two derivatives, then it follows that the right-hand side of~\eqref{proof_prop_AF_Pen_ATS_8d_equivalentce_between_Riccatti_and_local_martingale_constrained_Version} is closed in $C^2(\rr;\rr^{d+1})$; whence it is closed in the relative topology on $\hhh\subseteq C^{2}(\rr;\rr^{d+1})$.  
Thus, Assumption~(ii) holds.  
%

Lastly, since the loss function $\ell$, defined by
\begin{equation}
\ell(\varphi-\phi)\triangleq \int_{0}^{\infty}\int_{\beta \in \rrd} 
e^{-|u|^{\kappa}-\|\beta\|^{\kappa}}\left( 
\varphi(u,\beta)
-
\phi(u,\beta)
\right)^p dud\beta
\label{proof_prop_AF_Pen_ATS_9_continuityoflossfunction}
;
\end{equation}
is continuous on $\hhh$; then the conditions for Theorem~\eqref{thrm_Gamma_limit} are all met.  Therefore,~\eqref{thrm_AF_Pen_ATS_limittin_statement} holds.  

Next, (ii) will be verified.  Since second-order differential operators are continuous from $C^2(I\times \mmm \times \uuu)$ to $C^0(I\times \mmm \times \uuu)$, where the later is equipped with the convergence on compact topology, then functions
\begin{equation}
\begin{aligned}
\Phi\mapsto &\left|
c_0-\frac{\partial \Phi^0}{\partial u}(u) + \sum_{i=1}^d \gamma^i\Phi^i(u) - \frac1{2}\sum_{i,j=1}^d\alpha^{i,j}\Phi^i(u)^{\star}\Phi^j(u)
\right|^p\\
\Phi\mapsto &\left|
c_k-\frac{\partial \Phi^k}{\partial u}(u) + \sum_{i=1}^d \gamma^{k,i}\Phi^i(u) - \frac1{2}\sum_{i,j=1}^d\alpha_{k;i,j} \Phi^i(u)^{\star}\Phi^j(u)
\right|^p
,
\end{aligned}
\label{proof_prop_AF_Pen_ATS_10_continuity_in_phi_of_differential_operator}
\end{equation}
are continuous from $C^2(I\times \mmm \times \uuu)$ to $[0,\infty)$. Furthermore, since $W_w^{p,k}(I\times \mmm\times \uuu)$ is continuously embedded within $C^2(I\times \mmm\times \uuu)$ then the functions of~\eqref{proof_prop_AF_Pen_ATS_10_continuity_in_phi_of_differential_operator} are continuous from $W_w^{p,k}(I\times \mmm \times \uuu)$ to $[0,\infty)$.  The definition of the weight function in~\eqref{eq_growth_conditions} implies that, for every $\phi \in \hhh$ and every $\lambda \geq 2$, the integral is finite.  Thus, for every $\lambda \geq 2$, the map $\phi \mapsto \AF^{\lambda}(\phi)$ is continuous from $W_w^{p,k}(I\times \mmm \times \uuu)$ to $[0,\infty)$.  Furthermore, since the loss-function $\ell$ is continuous, then, for every $\lambda\geq 2$ the function $\Phi\mapsto \ell(\phi)+\AF^{\lambda}(\phi)$ is continuous.  Furthermore, since both $\ell$ and $\AF^{\lambda}$ are bounded below by $0$ and proper then so is $\ell(\varphi - \cdot) + \AF^{\lambda}(\cdot)$.  Lastly, since $\ell$ is coercive then by definition, for every $r\geq 0$, there exists a compact subset $K_r\subseteq \hhh$ satisfying
\begin{equation}
\left\{
\phi \in \hhh :\, \ell(\varphi-\phi)  \leq k
\right\}\subseteq K_r
.
\label{proof_prop_AF_Pen_ATS_11_coercivity_of_lambda}
\end{equation}
Therefore, the non-negativity of each $\AF^{\lambda}$ implies that for every $\lambda\geq 2$ and every $r\geq 0$,
\begin{equation}
\left\{
\phi \in \hhh :\, \ell(\varphi-\phi) + \AF^{\lambda}(\phi) \leq k
\right\}\subseteq 
\left\{
\phi \in \hhh :\, \ell(\varphi-\phi)  \leq k
\right\}\subseteq K_r
;
\label{proof_prop_AF_Pen_ATS_12_coercivity_of_LHS_objective_function}
\end{equation}
thus~\eqref{proof_prop_AF_Pen_ATS_12_coercivity_of_LHS_objective_function} implies that $\phi\mapsto \ell(\phi)+\AF^{\lambda}(\phi)$ is coercive.  Thus, for every $\lambda\geq 2$, the function $\phi\mapsto \ell(\phi)+\AF^{\lambda}(\phi)$ is lower semi-continuous, bounded-below, proper, and coercive on $\hhh$; thus by Weirestrass's Theorem, \cite[Theorem 2.2]{focardi2012gamma}, it admits a minimizer on $\hhh$.  
\end{proof}
Next, the arbitrage-free regularization of forward-rate curves will be considered using deep learning methods.  In this setting, it will be seen that~\eqref{AFReg} not only gains further computational tractability but, under a suitable initialization of the relevant deep feedforward neural networks, the loss-function described by Theorem~\ref{thrm_AF_Pen_ATS} also allows further simplification.  
\subsection{A Deep Learning Approach to Arbitrage-Free Regularization}
The flexibility of feedforward artificial neural networks (ANNs), as described in the universal approximation theorems of \cite{hornik1990universal,Cybenko}, makes the collection of ANNs a well-suited class of alternative models for the arbitrage-free regularization problem.  In the context of this paper an ANN is any function from $\rrd$ to $\rrd$ of the form
\begin{equation}
W_{N+1} \circ 
\rho\bullet W_N \circ \dots \rho\bullet W_1
\label{eq_ANN_description}
,
\end{equation}
where $\{W_i\}_{i=1}^{N+1}$ are affine functions from $\rr^{d_i}$ to $\rr^{d_{i+1}}$ where $d_1=d=d_{N+1}$, $\rho$ is a continuous activation-function, and $\bullet$ denotes component-wise composition.  
However, in order to maintain analytic tractability, it will be required that the class of alternative models still be of affine type; thus, in our analysis, $\hhh$ will consist of all functions in $W^{p,k}_w(I\times \mmm \times \uuu)$ of the form
\begin{equation}
\phi(t,\beta,u) = 
\underline{\beta}^T\left(W_{N+1} \circ 
\rho\bullet W_N \circ \dots \circ \rho\bullet W_1 (u+t)
\right)
\label{eq_alternative_FRC_models}
,
\end{equation}
where $\underline{\beta}_1=1$ and $\underline{\beta}_{i+1}=\beta^i$ for all $i>1$.  

It has been shown in \cite{rahimi2008random,saxe2011random,choromanska2015loss,louart2018random}, amongst others, that if a network is appropriately designed, then only training the final layer and suitably initializing the matrices $W_N,\dots,W_1$ performs comparably well to networks with all the layers trained.  This phenomenon has been observed in numerous numerical studies, such as \cite{jaeger2004harnessing}, where the entries of the matrices $W_N,\dots,W_1$ are chosen entirely randomly.  This practice has also become fundamental to feasible implementations of recurrent neural network (RNN) theory and reservoir computing, as studied in \cite{gelenbe1989random,gelenbe1993learning,gelenbe1993dynamical,bakirciouglu2000survey,yin2017single}, where training speed becomes a key factor in determining the feasibility of the RNN and reservoir computing paradigms.  

Following this practice, the hypothesis class of alternative factor-models to be considered in the arbitrage-free regularization problem, effectively reduces from~\eqref{eq_alternative_FRC_models} to
\begin{equation}
\begin{aligned}
\phi(t,\beta,u) = &
\underline{\beta}^T Wf(u+t)
,
\end{aligned}
\label{eq_alternative_FRC_models_redux}
\end{equation}
where $W\triangleq W_{N+1}$, $f\triangleq \rho\bullet W_N \circ \dots \circ \rho\bullet W_1 (u+t)$, and $W_1,\dots,W_N$ are initialized either randomly or otherwise.  In this implementation, $F$ is initialized implicitly through the following optimization problem
\begin{equation}
\tilde{f} \in \argmin \, \int_0^{\tilde{T}} \left(f_i(u)-\phi_i(u)\right)^p e^{-|u|^{\kappa}}du
,
\label{eq_optimization_problem_initializing_reservoirF}
\end{equation}
where the $\argmin$ is taken over all deep feedforward neural networks, as in~\eqref{eq_alternative_FRC_models}, with fixed depth $N$, fixed height equal (dimension of each $W_i$), fixed $\tilde{T}>0$, input $\rr$, and output in $\rr^3$.  
The network $f$ in~\eqref{eq_alternative_FRC_models_redux} is then defined by removing the final layer from $\tilde{f}$.  The removal of the final layer both increases the dimension of the output of $f$, which provides more flexibility when optimizing~\eqref{AFReg} over \eqref{eq_alternative_FRC_models_redux}, and simultaneously ensures that $f$ is capable of describing the original factor-model of ATS type.

Furthermore, under the simplified structure provided by the hypothesis class~\eqref{eq_alternative_FRC_models_redux}, the quantity $\Lambda_t^u(\phi)$ in~\eqref{thrm_AF_Pen_ATS_definition_AF_penalty}, defining the arbitrage-penalty, can be further simplified.  A brief computation shows for any $\phi$ as in~\eqref{eq_alternative_FRC_models_redux}, $\Lambda_t^u(\phi)$ simplifies to
	\begin{equation}
\begin{aligned}
\Lambda^u(W)\triangleq &\left|
W^0f(u)-W^0f(0) + \sum_{i=1}^d \gamma^iW^i F(u) - \frac1{2}\sum_{i,j=1}^d\alpha^{i,j} F(u)^{\star}(W^i)^{\star}W^j F(u)
\right|^p\\
+\sum_{k=1}^d 
&\left|
W^kf(u)-W^kf(0) + \sum_{i=1}^d \gamma^{k,i}W^i F(u) -
\frac1{2}\sum_{i,j=1}^d\alpha_{k;i,j} F(u)^{\star}(W^i)^{\star}W^j F(u)
\right|^p
,
\end{aligned}
\label{eq_AF_penalty_reduced_for_Randomized_ANN_setting}
\end{equation}
where $F(u)\triangleq \int_0^u f(s)ds$ with the integration is defined component-wise and where $W^{\star}$ denotes the transpose of $W$ and $W^i$ denotes the $i^{th}$ row of the matrix $W$.  

\subsection{Numerical Implementations}
The performance of the arbitrage-free regularization methodology will now be applied to two factor-models, of affine type, and its performance will be evaluated numerically.  The first factor-model is the commonly used {dynamic Nelson-Siegel} model of \cite{YieldDbold} and the second is a machine learning extension of the classical PCA approach to term-structure modeling.  The performance of the arbitrage-free regularization, for each model, will be benchmarked against both the original factor-models and against the HJM-extension of the Vasi\v{c}ek model.  The Vasi\v{c}ek model is a natural benchmark since, as shown in \cite{bjork1999interest}, it is consistent with a low-dimensional factor-model.  Therefore, each of the factor-models contains roughly the same number of driving factors; which ensures that the comparisons are considered fair.  

The data-set for this implementation consists of German bond data for 31 maturities with observations obtained on 1273 trading days from January $4^{th}$ 2010 to December $30^{th}$ 2014.  As is common practice in machine learning, further details of our code as the implementation itself can be found on 
\cite{GITHUBDeepAFReg}.

As described in~\eqref{eq_alternative_FRC_models_redux}-\eqref{eq_AF_penalty_reduced_for_Randomized_ANN_setting}, the solution to the arbitrage-free regularization~\eqref{AFReg}, will be numerically approximated using initialized deep feed-forward neural networks.  The initialization network $f$, of~\eqref{eq_alternative_FRC_models_redux}, is selected to have fixed depth $N=5$, fixed height $d=d_i=10^2$ for all but the first and last layers, and its weights is learned using the ADAM algorithm of \cite{kingma2014adam}.  
The meta-parameters $p=2$ and $\kappa=1$ are chosen empirically, and the parameters of the Ornstein-Uhlenbeck process are estimated following the maximum-likelihood procedure described in \cite{MeucciRiskandAssetAllocation}.  Once the model parameters have been learned, and the factor-model optimizing~\eqref{AFReg} has been learned, the one day ahead predictions of the stochastic factors are obtained through Kalman filter estimates, see \cite{AlainFilter} for details on the Kalman filter.  These predictions are then compared to the realized next-day bond prices.

Before turning to the model's performance, each factor-model will be briefly overviewed.  
\subsubsection{Model 1: The Dynamic Nelson-Siegel Model}
The Nelson-Siegel family is a low-dimensional family of forward-rate curve models used by various countries' central banks to produce forward-rate or yield curves.  As discussed in \cite{CarmonaR}, Finland, Italy, and Spain are such examples with other countries such as Canada, Belgium, and France relying on a slight extension of this model.  The Nelson-Siegel model's popularity is largely due to its interpretable factors and satisfactory empirical performance.  It is defined by
\begin{equation}
\varphi(t,\beta,u)\triangleq \beta^1 + \beta^2 e^{-(u+t)\tau} + \beta^3 (u+t)e^{-(u+t)\tau},
\label{eq_Nelson_Siegel_Definition}
\end{equation}
where, as discussed in \cite{YieldDbold}, the first factor represents the long-term level of the forward-rate curve, the second represents its shape, the third represents its curvature, and $\tau$ is a shape parameter; typically kept fixed.  

Since market conditions are continually changing, the Nelson-Siegel model is typically extended from a static state-space model to a dynamic state-space model, by replacing the static choice of $\beta$ with a three-dimensional Ornstein-Uhlenbeck process and fixing the shape parameter $\tau>0$, as in \cite{YieldDbold}.  This is the \textit{dynamic Nelson-Siegel} model.  However, as demonstrated in \cite{filipovic2001consistency}, the dynamic Nelson-Siegel model does not admit an equivalent measure to $\pp$, which makes the entire bond market simultaneously into local martingales.  It was then shown in \cite{AFNS} that a specific additive perturbation of the Nelson-Siegel family circumvents this problem, but empirically this is observed to come at the cost of reduced predictive accuracy.  In our implementation, the parameters of the Ornstein-Uhlenbeck process driving $\beta_t^i$ will be estimated using the maximum likelihood method described in \cite{MeucciRiskandAssetAllocation}.  
%

\subsubsection{Model 2: An Algorithmic Affine-Term Structure Model}
The dynamic Nelson-Siegel model's shape has been developed through practitioner experience.  The second factor-model considered here will be of a different type, as its factors will be learned algorithmically.  Similarly to~\eqref{eq_Nelson_Siegel_Definition}, first a static three-factor-model for the forward-rate curve of the form
\begin{equation*}
\varphi(t,\beta,u)\triangleq \beta^0 + \sum_{i=1}^{3}\beta^i \phi_i(u+t),
\end{equation*}
where each $\phi_1,\dots,\phi_3$ are the first three principal components of the forward-rate curve; calibrated on the first 100 days of data.  Then, a time-series for the $\beta^i$ parameters is generated, again using the first 100 days of data, where on each day $\beta_0,\dots,\beta_3$ optimizes a cross-validated LASSO regression problem, as introduced in \cite{tibshirani1996regression}, with the response variables being the points observed on that day's observed forward-rate curve.  Finally, analogously to the dynamic Nelson-Siegel model, an $\rr^4$-valued of the Ornstein-Uhlenbeck process is calibrated to the time-series of the $\beta^i$ parameters using the maximum likelihood methodology outlined in \cite{MeucciRiskandAssetAllocation}.  This defines an affine-term structure model, whose factor-model $\varphi$ represents a linear combination the principal components of the forward-rate curve and an intercept, and whose stochastic factors $\beta_t^i$ are calibrated to solutions to daily cross-validated LASSO regularized regression problems.  We will denote this model by dPCA and its arbitrage-free regularization by AF-Reg(dPCA)

\subsubsection{Results}
Tables~\ref{tab_AFReg_NS_Compare_short},~\ref{tab_AFReg_NS_Compare_mid},~\ref{tab_AFReg_NS_Compare_long},~\ref{tab_AFReg_NS_Compare_30year_long} compare the predictive performance of the Vasi\v{c}ek (Vasi\v{c}ek), dPCA, AF-Reg(dPCA), the dynamic Nelson-Siegel Model (dNS), the arbitrage-free Nelson-Siegel model of \cite{AFNS} (AFNS), and the arbitrage-free regularization of the dynamic Nelson-Siegel Model (AF-Reg(NS)).  The predictive quality is quantified by the estimated mean-squared errors when making one day-ahead predictions of the bond price for each maturity, for all but the first days in our data-set.  The lowest estimated mean-squared error recorded are highlighted using bold font and the second lowest estimated mean-squared error on each maturity are emphasized using italics.  
\begin{table}[H]
	\centering
	\begin{tabular}{rrrrrr}
		\hline
	Model $\backslash$ Maturity	& 0.5 & 1 & 2 & 3 & 4 \\ 
		\hline
		Vasi\v{c}ek & 3.155e-01 & 4.323e-01 & 3.622e-01 & 1.950e-01 & \textbf{5.730e-02} \\ 
		dPCA & \textit{2.526e-01} & 4.349e-01 & 4.176e-01 & 2.526e-01 & 9.261e-02 \\ 
		AF-Reg(dPCA)& 8.066e-01 & 6.943e-01 & 5.110e-01 & 2.755e-01 & 9.588e-02 \\ 
		NS & 4.513e-02 & \textit{1.479e-01} & \textit{2.134e-01} & \textit{1.477e-01} & \textit{5.968e-02} \\ 
		AFNS & 4.513e-02 & \textit{1.479e-01} & \textit{2.134e-01} & \textit{1.477e-01} & \textit{5.968e-02} \\ 
		AF-Reg(NS) & \textbf{2.903e-02} & \textbf{9.514e-02} & \textbf{1.601e-01} & \textbf{1.235e-01} & 6.482e-02 \\ 
		\hline
	Model $\backslash$ Maturity	& 5 & 6 & 7 & 8 & 9\\
		\hline
		Vasi\v{c}ek & \textbf{7.735e-03} & \textbf{1.996e-04} & 1.024e-03 & 1.480e-03 & 1.348e-03 \\ 
		dPCA & 2.193e-02 & 3.326e-03 & \textbf{3.119e-04} & \textbf{1.897e-05} & \textbf{8.097e-07} \\ 
		AF-Reg(dPCA)& 2.221e-02 & 3.340e-03 & \textit{3.123e-04} & \textit{1.898e-05} & \textit{8.099e-07} \\
		NS & \textit{1.972e-02} & \textit{8.313e-03} & 5.323e-03 & 3.925e-03 & 2.998e-03 \\ 
		AFNS & \textit{1.972e-02} & \textit{8.313e-03} & 5.323e-03 & 3.925e-03 & 2.998e-03 \\ 
		AF-Reg(NS) & 3.579e-02 & 2.236e-02 & 1.523e-02 & 1.050e-02 & 7.308e-03 \\ 
		\hline
	\end{tabular}
	\caption{(Short): MSE Comparisons for 1-day ahead bond-price predictions.} 
	\label{tab_AFReg_NS_Compare_short}
\end{table}
\begin{table}[H]
	\centering
	\begin{tabular}{rrrrrr}
		\hline
	Model $\backslash$ Maturity	& 10 & 11 & 12 & 13 & 14 \\ 
		\hline
		Vasi\v{c}ek & 1.108e-03 & 9.002e-04 & 7.382e-04 & 6.125e-04 & 5.135e-04 \\ 
		dPCA & \textbf{2.578e-08} & \textbf{6.328e-10} & \textbf{1.433e-11} & \textbf{2.607e-13} & \textbf{4.179e-15} \\ 
		AF-Reg(dPCA)& \textit{2.579e-08} & \textit{6.328e-10} & \textit{1.433e-11} & \textit{2.607e-13} & \textit{4.179e-15} \\ 
		NS & 2.381e-03 & 1.969e-03 & 1.686e-03 & 1.484e-03 & 1.337e-03 \\ 
		AFNS & 2.381e-03 & 1.969e-03 & 1.686e-03 & 1.484e-03 & 1.337e-03 \\ 
		AF-Reg(NS) & 5.215e-03 & 3.827e-03 & 2.885e-03 & 2.229e-03 & 1.761e-03 \\ 
		\hline
	Model $\backslash$ Maturity	& 15 & 16 & 17 & 18 & 19\\
		\hline
		Vasi\v{c}ek & 4.342e-04 & 3.698e-04 & 3.169e-04 & 2.729e-04 & 2.360e-04 \\ 
		dPCA & \textbf{6.714e-17} & \textbf{9.566e-19} & \textbf{1.426e-20} & \textbf{1.819e-22} & \textbf{2.749e-24} \\ 
		AF-Reg(dPCA)& \textit{6.714e-17} & \textit{9.566e-19} & \textit{1.426e-20} & \textit{1.818e-22} & \textit{2.746e-24} \\ 
		NS & 1.225e-03 & 1.138e-03 & 1.069e-03 & 1.012e-03 & 9.639e-04 \\ 
		AFNS & 1.225e-03 & 1.138e-03 & 1.069e-03 & 1.012e-03 & 9.639e-04 \\ 
		AF-Reg(NS) & 1.422e-03 & 1.171e-03 & 9.831e-04 & 8.406e-04 & 7.316e-04 \\ 
		\hline
	\end{tabular}
	\caption{(Mid): MSE Comparisons for 1-day ahead bond-price predictions.} 
	\label{tab_AFReg_NS_Compare_mid}
\end{table}
\begin{table}[H]
	\centering
	\begin{tabular}{rrrrrr}
		\hline
	Model $\backslash$ Maturity	& 20 & 21 & 22 & 23 & 24 \\ 
		\hline
		Vasi\v{c}ek & 2.049e-04 & 1.784e-04 & 1.558e-04 & 1.364e-04 & 1.196e-04 \\ 
		dPCA & \textbf{3.816e-26} & \textbf{5.254e-28} & \textbf{8.047e-30} & \textbf{9.958e-32} & \textbf{1.336e-33} \\ 
		AF-Reg(dPCA) & \textit{3.781e-26 }& \textit{4.847e-28} & \textit{3.015e-30} & \textit{2.684e-30} & \textit{1.452e-29} \\
		NS & 9.228e-04 & 8.866e-04 & 8.542e-04 & 8.247e-04 & 7.976e-04 \\ 
		AFNS & 9.228e-04 & 8.866e-04 & 8.542e-04 & 8.247e-04 & 7.976e-04 \\ 
		AF-Reg(NS) & 6.480e-04 & 5.838e-04 & 5.349e-04 & 4.984e-04 & 4.814e-04 \\ 
		\hline
	Model $\backslash$ Maturity	& 25 & 26 & 27 & 28 & 29\\
		\hline
		Vasi\v{c}ek & 1.051e-04 & 9.254e-05 & 8.160e-05 & 7.205e-05 & 6.371e-05 \\ 
		dPCA & \textbf{2.067e-35} & \textbf{2.814e-37} & \textbf{3.639e-39} & \textbf{5.371e-41} & \textbf{7.459e-43} \\ 
		AF-Reg(dPCA) & \textit{9.846e-29} & \textit{1.102e-27} & \textit{2.108e-26} & \textit{6.986e-25} & \textit{3.979e-23} \\ 
		NS & 7.722e-04 & 7.484e-04 & 7.257e-04 & 7.041e-04 & 6.835e-04 \\ 
		AFNS & 7.722e-04 & 7.484e-04 & 7.257e-04 & 7.041e-04 & 6.835e-04 \\ 
		AF-Reg(NS) & 4.911e-04 & 5.288e-04 & 6.011e-04 & 7.214e-04 & 9.138e-04 \\ 
		\hline
	\end{tabular}
	\caption{(Long): MSE Comparisons for 1-day ahead bond-price predictions.} 
	\label{tab_AFReg_NS_Compare_long}
\end{table}
\begin{table}[H]
	\centering
	\begin{tabular}{rr}
		\hline
		& 30 \\ 
		\hline
		Vasi\v{c}ek & 6.371e-05 \\ 
		dPCA & \textbf{7.459e-43} \\ 
		AF-Reg(dPCA) &\textit{ 3.979e-23} \\ 
		NS & 6.835e-04 \\ 
		AFNS & 6.835e-04 \\ 
		AF-Reg(NS) & 9.138e-04 \\ 
		\hline
	\end{tabular}
	\caption{(30 Year): MSE Comparisons for 1-day ahead bond-price predictions.} 
	\label{tab_AFReg_NS_Compare_30year_long}
\end{table}
Our numerical implementations highlight a few key facts about the arbitrage-free regularization methodology.  First, for nearly every maturity, the empirical performance of the arbitrage-free regularization of a factor-model is comparable to the original factor-model.  An analogous phenomenon was observed in \cite{devin2010finite} when projecting infinite-dimensional arbitrage-free HJM models onto the finite-dimensional manifold of Nelson-Siegel curves.  Therefore, correcting for arbitrage does not come at a significant predictive cost.  However, it does come with the benefit of making the model theoretically sound and compatible with the techniques of arbitrage-pricing theory.  

Second, since~\eqref{AFReg} incorporates an additional constraint into the modeling procedure the arbitrage-free regularization of a factor-model has a reduction in performance as compared to the initial factor-model.  This phenomenon, has also been observed empirically in \cite{AFNS} for the arbitrage-free Nelson-Siegel correction of the dynamic Nelson-Siegel model.  Therefore, one should not expect to improve on the predictive performance of the initial factor-model by correcting for the existence of arbitrage.  

Third, the empirical performance of AF-Reg(dPCA) was significantly better than the empirical performance of the other arbitrage-free models, namely AFNS, AF-Reg(NS), and the Vasi\v{c}ek model, across nearly all maturities.  This was especially true for mid and long maturity zero-coupon bonds.   Moreover, analogous to our second point, the performance of AF-Reg(dPCA) and dPCA were comparable.  Similarly, for most matirities, the empirical performance of the AFNS, dNS, and AF-Reg(NS) models were all similar and notably lower than the performance of the AF-Reg(dPCA), dPCA, and Vasi\v{c}ek models.  This emphasizes the fact that arbitrage-free regularization methodology produces performant models only if the original model itself produces accurate predictions.  Therefore, it is up to the modeller to make an appropriate choice of model; however the methodology used to develop dPCA and AF-Reg(dPCA) could be used as a generic starting point
.   

Care must be taken in selecting an appropriate factor-model.  However, since the arbitrage-free regularization methodology applies to nearly any factor-model, one may use any methodology to produce a accurate factor-model and then appeal to arbitrage-free regularization to make it theoretically consistent at a small cost in performance.  This opens the possibility to applying machine learning models, such as dPCA to finance without the worry that they are, or are not, arbitrage-free since their arbitrage-free regularization is well-defined.  
Furthermore, the flexibility of deep feed-forward neural networks allows one to implement~\eqref{AFReg}.  This problem may have been previously unfeasible before the advent of this generic computational method.  

We summarize the contributions made in by this article.  
\section{Conclusion}
This paper introduced a novel model-selection problem and approach was introduced for learning arbitrage-free factor-models in a generalized HJM-type framework, as formulated by~\eqref{AF}.  By exploiting the structure of our generalized HJM-type models, a generalization of the consistency condition of \cite{filipovic2001consistency} and the drift condition of \cite{heath1992bond} was obtained in Proposition~\ref{prop_NA_characterization}, from which the alternative computationally-tractable model selection~\eqref{AFReg} could be formulated.  Using techniques from the theory of $\Gamma$-convergence, introduced by \cite{de1980gamma}, Theorem~\ref{thrm_Gamma_limit} showed that the optimizers of the tractable model selection~\eqref{AFReg} converge to an optimizer of the computationally intractable model-selection~\eqref{AF}.  

Returning to the setting of term-structure models, Theorem~\ref{thrm_AF_Pen_ATS} affirmed that the conditions for Theorem~\ref{thrm_Gamma_limit} are indeed met in the affine term-structure setting, and a reduced form of~\eqref{AFReg} was obtained.  
Following Theorem~\ref{thrm_AF_Pen_ATS}, a deep-learning approach to optimization~\eqref{AFReg} was taken, and numerical experiments were considered comparing the arbitrage-free regularization approach applied to the dynamic Nelson-Siegel and an algorithmically generated model to other comparable models.  It was found that, in general, the arbitrage-free regularization method did not come at a significant predictive cost in making 1-day ahead bond price predictions.  

Therefore, the arbitrage-free regularization methodology allows one to utilize algorithmic models, such as dPCA, in a way which is consistent with the efficient market hypothesis and with arbitrage-pricing theory without a notable loss in performance.  In future research, the authors plan to evaluate the performance of algorithmic factors model in other types of large financial market, in a way which is consistent with classical arbitrage pricing theory, using the arbitrage-free regularization methodology.  

The authors would like to thank the ETH Z\"{u}rich Foundation as well as by the Natural Sciences and Engineering Research Council of Canada (NSERC) for the funding required to pursue this research project.  The authors would equally like to thank Alina Stancu for many helpful discussions.
\renewcommand{\thepage}{}
\bibliographystyle{abbrvnat}
\bibliography{References}
\printindex
\end{document}